\documentclass[twocolumn,trackchanges]{aastex701}

\usepackage{comment}

\shorttitle{From blue to red spirals via RPS}
\shortauthors{Lora et al.}

\begin{document}
\title[From blue to red spirals: Slow galaxy transformation via RPS in TNG-50] {From blue to red spirals: \\
Slow galaxy transformation via ram pressure stripping in TNG-50}

\author[0000-0003-3588-5235]{V. Lora}
\affiliation{Instituto de Ciencias Nucleares (UNAM), Ap. 70-264, C.P. 04510, Mexico City, Mexico}
\email[show]{v.lora@nucleares.unam.mx} 

\author{J. I. González-Carbajal}
\affiliation{Instituto de Ciencias Nucleares (UNAM), Ap. 70-264, C.P. 04510, Mexico City, Mexico}
\email{v.lora@nucleares.unam.mx} 

\author{A. Pasquali}
\affiliation{Astronomisches Rechen-Institut, Zentrum f\"ur Astronomie der Universit\"at Heidelberg, M\"onchhofstr. 12--14, 69120 Heidelberg, Germany}
\email{v.lora@nucleares.unam.mx} 

\author{A. Marasco}
\affiliation{INAF - Padova Astronomical Observatory, Vicolo dell’Osservatorio 5, 35122 Padova, Italy}
\email{v.lora@nucleares.unam.mx} 

\author{J. Fritz}
\affiliation{Instituto de Radioastronomía y Astrofísica (UNAM), Antigua Carretera a P\'atzcuaro 8701, 58089 Morelia, Mexico}
\email{v.lora@nucleares.unam.mx} 

\author{E. K. Grebel}
\affiliation{Astronomisches Rechen-Institut, Zentrum f\"ur Astronomie der Universit\"at Heidelberg, M\"onchhofstr. 12--14, 69120 Heidelberg, Germany}
\email{v.lora@nucleares.unam.mx} 

\author{J. P. Ortíz}
\affiliation{Instituto de Astronomía (UNAM),
Ap. 70-264, C.P. 04510, Mexico City, Mexico}
\email{v.lora@nucleares.unam.mx} 

\author{S. Estrada-Dorado}
\affiliation{Instituto de Ciencias Nucleares (UNAM), Ap. 70-264, C.P. 04510, Mexico City, Mexico}
\email{v.lora@nucleares.unam.mx}

%%==================================%%
%%            abstract              %%
%%==================================%%
\begin{abstract}
Late-type galaxies lose gas through ram-pressure stripping (RPS) after falling into a massive halo. Because this mechanism primarily removes the gaseous component while leaving the stellar disk largely undisturbed, it provides a pathway for quenching star formation without immediate morphological transformation. While RPS is well established in galaxy clusters, galaxy evolution in low-mass groups is often attributed to mergers, leaving open the question of whether RPS alone can drive the transition from blue, star-forming spirals to quenched systems in these environments. We use the high-resolution cosmological simulation TNG-50 to investigate the evolution of blue spiral galaxies, after their infall into group-scale halos. We excluded systems undergoing significant mergers, thus isolating the effect of RPS. We find that RPS in low-mass groups (M$_{group}<10^{14.5}$ M$_{\odot}$) can efficiently quench star formation while preserving the stellar disk structure. The transformation is gradual, with quenching timescales $\gtrsim6$ Gyr after infall, longer than the $\sim4$ Gyr typically associated with merger-driven evolution. The resulting galaxies are predominantly red, anemic spirals rather than fully transformed S0 systems, indicating that gas removal alone is insufficient to produce complete morphological transformation. Our results show that RPS in group environments can generate long-lived quenched spirals which might represent an intermediate evolutionary pathway preceding the formation of lenticular galaxies.
\end{abstract}

%%================================%%
%%          KeyWords              %%
%%================================%%

%%================================%%
%%       Introduction             %%
%%================================%%
\section{Introduction}

Understanding how galaxies evolve across different environments remains a central challenge. Galaxy evolution is driven by a combination of internal processes (including star formation, stellar and AGN feedback, etc.), and external environmental mechanisms such as mergers, tidal interactions, and hydrodynamical effects \citep{peng:10,pasquali:10,pasquali:12,galazzi:21}.

The influence of environment is clearly imprinted in the observed morphology–density relation: dense regions such as groups and clusters host a higher fraction of red, quenched, early-type galaxies, while field regions are dominated by blue, star-forming spirals \citep{dressler:80,butcher:84}. This environmental dependence extends to stellar populations, gas fractions, and star formation activity, suggesting that dense environments actively suppress star formation.

A key hydrodynamical mechanism driving environmental transformation is ram pressure stripping (RPS) \citep{gunn:72}. As a galaxy moves through the hot intracluster (ICM) or intragroup medium, it experiences a pressure proportional to the medium gas density, and to the square of the galaxy's velocity relative to the medium. When this pressure exceeds the gravitational restoring force of the disk, cold gas is removed, leading to a rapid decline in star formation. Hydrodynamical simulations demonstrate that RPS removes gas in a progressive, outside-in manner \citep{roediger:07,tonnesen:09,bekki:14}, consistent with truncated star-forming disks observed in cluster spirals.

Observationally, ram-pressure stripping (RPS) is most strikingly associated with one-sided gas tails, as seen in the so-called jellyfish galaxies (e.g., \cite{poggianti:19}). Because RPS primarily affects the gaseous component, the stellar disk typically remains dynamically undisturbed, preserving rotational support even as star formation is suppressed \citep{boselli:08,cortese:11}. 
However, jellyfish galaxies likely represent only the most extreme manifestations of RPS. A more global view is provided by spatially resolved HI studies in nearby clusters, such as the seminal work of \citeauthor{cayette:90} \citeyear{cayette:90}, which revealed a systematic truncation of gaseous disks in Virgo spirals, with galaxies in the cluster core retaining only $\sim20$\% of their HI mass when compared with field galaxies. Subsequent multi-wavelength studies, including HI and X-ray mapping, have reinforced this picture, showing both the progressive removal of cold gas and its interaction with the hot ICM (e.g., \cite{chung:09,wang:21}). 

Simulations further illustrate the temporal evolution of RPS, from early phases characterized by extended gaseous tails to later stages in which galaxies are largely stripped of their gas while the ambient medium relaxes on Gyr timescales (e.g., \citep{vijayaraghavan:15}).

Both theoretical and observational studies indicate that RPS can quench star formation on relatively short timescales ($\sim1$--$2$~Gyr; referring to the time elapsed since a galaxy has fallen into the host halo) in cluster environments \citep{wetzel:13,fillingham:16,oman:16}. In such environments, stripping contributes significantly to the formation of lenticular (S0) galaxies, with models and observations suggesting that a substantial fraction of cluster S0s may originate from rapidly stripped spiral progenitors \citep{boselli:08,deeley:21}. Semi-analytic models incorporating environmental gas removal further show that ram-pressure and related stripping mechanisms help reproduce the observed increase in red, early-type galaxy fractions with environmental density \citep{bosch:08,guo:11,cora:18,marasco:26}.

A complementary evolutionary pathway involves gravitational interactions and mergers, which are more efficient in low-density environments where galaxy relative velocities are lower. Tidal encounters and mergers can remove gas, induce central starbursts, and redistribute angular momentum, often increasing central concentration and reducing rotational support \citep{querejeta:17,coccato:22}. Simulations suggest that merger-driven S0 formation proceeds on longer timescales ($\gtrsim4$~Gyr) and may dominate in low-density regions \citep{deeley:21}.

In addition to these gravitational processes, several mechanisms acting on the gaseous component may contribute to galaxy transformation. These include viscous stripping \citep{nulsen:82}, thermal evaporation due to heat conduction from a hot ambient medium \citep{cowie:77}, and starvation (or strangulation), in which the supply of fresh gas accreting onto galaxies is suppressed \citep{larson:80,bekki:02}. In the latter scenario, quenching does not result solely from the consumption of the existing cold gas reservoir, but also from the lack of continued gas infall required to sustain star formation over cosmological timescales; for instance, galaxies such as the Milky Way likely require ongoing accretion at a rate of order $\sim1,M_{\odot},\mathrm{yr}^{-1}$ to maintain their star formation over a Hubble time. While RPS is generally expected to be the dominant mechanism in dense environments, these processes likely act together; starvation can limit gas replenishment prior to or during infall, while viscous stripping and thermal evaporation can further weaken the gaseous component, enhancing the efficiency of subsequent stripping.

% REFEREE COMMENT 4 (TIDES)
\textbf{
Gravitational environmental effects may also contribute to the evolution of satellite galaxies. In addition to mergers and close encounters, the tidal field of the host halo can perturb both the stellar and gaseous components, leading to mass loss and quenching over extended timescales. Semi-analytic models have shown that environmentally driven tidal stripping can reproduce several observed trends in the color bimodality of satellite galaxies in groups and clusters \citep{weinmann:10}. In practice, multiple environmental mechanisms are expected to operate simultaneously, making it challenging to isolate their relative contributions in both observations and cosmological simulations.}

Although RPS is most efficient in massive clusters, 
ram pressure could also be important in lower-mass halos such as galaxy groups \citep{bahe:15}. In these environments, satellite galaxies may experience extended gas depletion over several Gyr, consistent with delayed-then-rapid quenching scenarios \citep{wetzel:13}. High-resolution cosmological simulations such as TNG50 reveal systematic outside-in shrinking of star-forming gas disks following infall, supporting the concept of slow environmental stripping in group-scale halos \citep{RohrJellyfishTNG502023,zinger:24}.

%%================================%%
%%          Figure 1              %%
%      Merger and Flybys          %%
%%================================%%
\begin{figure}[t!]
    \centering
    {\includegraphics[width=0.38\textwidth]{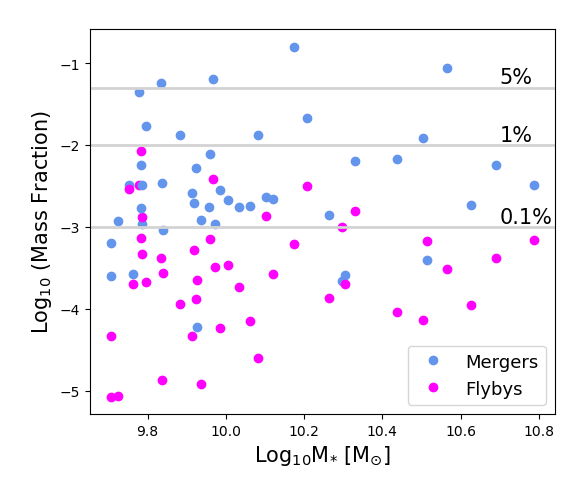}}\\
    {\includegraphics[width=0.38\textwidth]{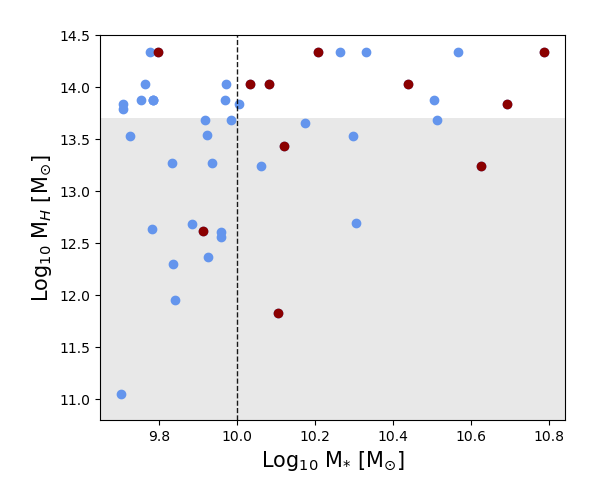}}
    \caption{The top panel shows the fraction of stellar mass acquired during mergers (blue dots), and flybys (magenta dots) since $z=1$, as a function of the subhalo stellar mass (at $z=0$). The three horizontal lines indicate fractions of $5\%$, $1\%$, and  $0.1\%$ of the total stellar mass in mergers and in flybys.
    The bottom panel shows the total mass of the halo (group) to which each subhalo belongs, as a function of the stellar mass of the subhalo. The shaded area divides the massive halos (M$_{H}>10^{13.5}$ M$_{\odot}$) from the low mass halos. The vertical line divides the high mass subhalos (M$_{*}>10^{10}$ M$_{\odot}$) from the low mass ones. \textbf{Those subhalos which belong to our red sample are highlighted in red, see Section \ref{subsec:BRG}.} }
    \label{Fig1}
\end{figure}

In parallel, observations have identified galaxies that challenge the simple blue/red dichotomy: red spirals (or “anemic spirals” \citeauthor{bergh:76} \citeyear{bergh:76}). These systems retain spiral morphology but exhibit red colors and low star formation rates, indicative of significant gas depletion \citep{poggiant:99,masters:10}. Red spirals have been proposed as transitional systems that are quenched without undergoing major structural disruption, potentially preceding transformation into S0 galaxies \citep{wolf:09}. Structural analyses reveal similarities between red spirals and S0s in Sérsic index, concentration, and stellar surface density \citep{cui:24}, while their enhanced frequency in intermediate and low density environments suggests that their quenching pathway may differ from classical cluster-driven transformation \citep{masters:10,RobertsRedSpiralsAandA2022}.

%%================================%%
%%          Figure 2              %%
%         SDSS EVOLUTION          %%
%%================================%%
\begin{figure*}[t!]
    \centering
    {\includegraphics[width=0.495\textwidth]{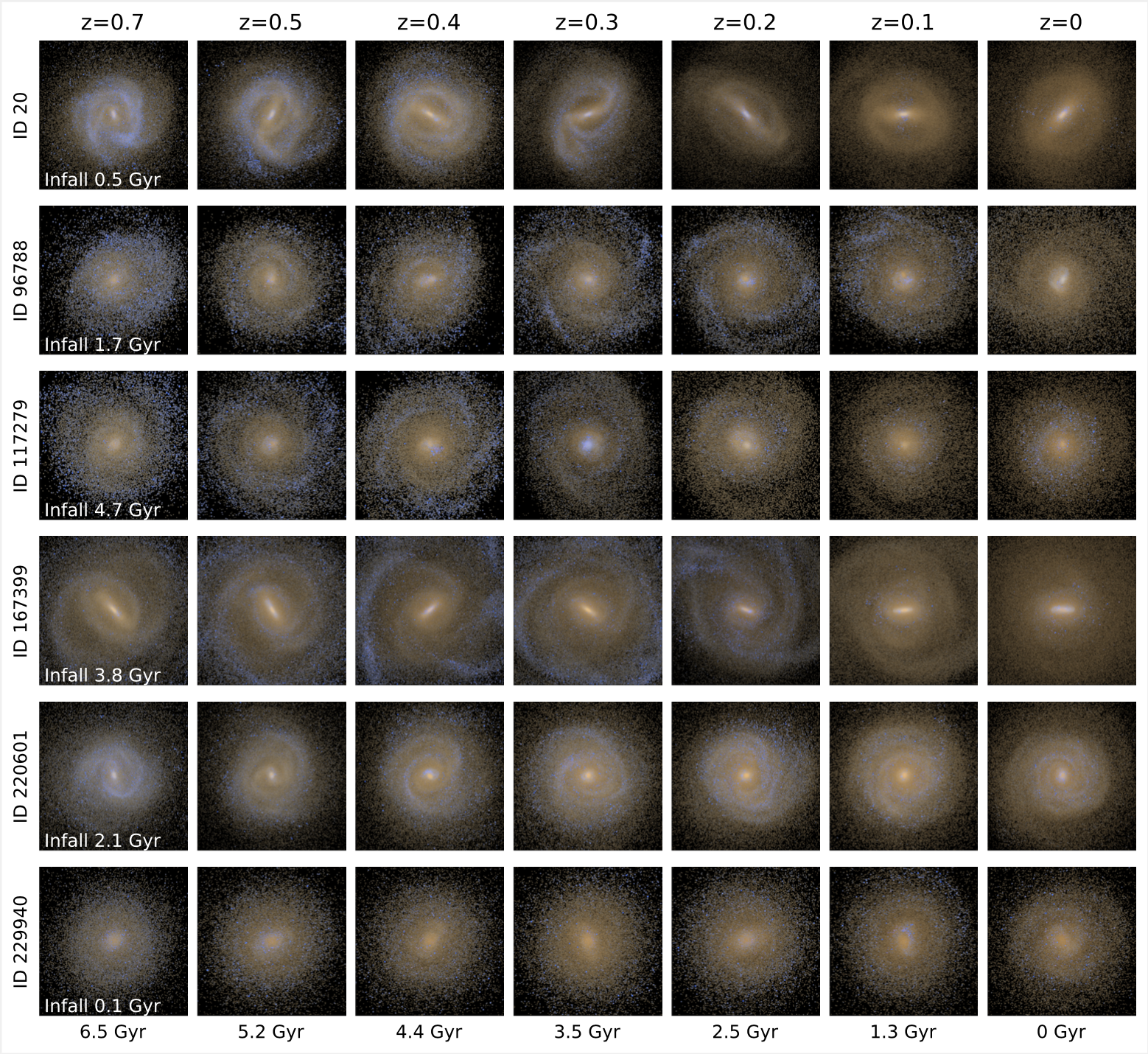}}
     {\includegraphics[width=0.495\textwidth]{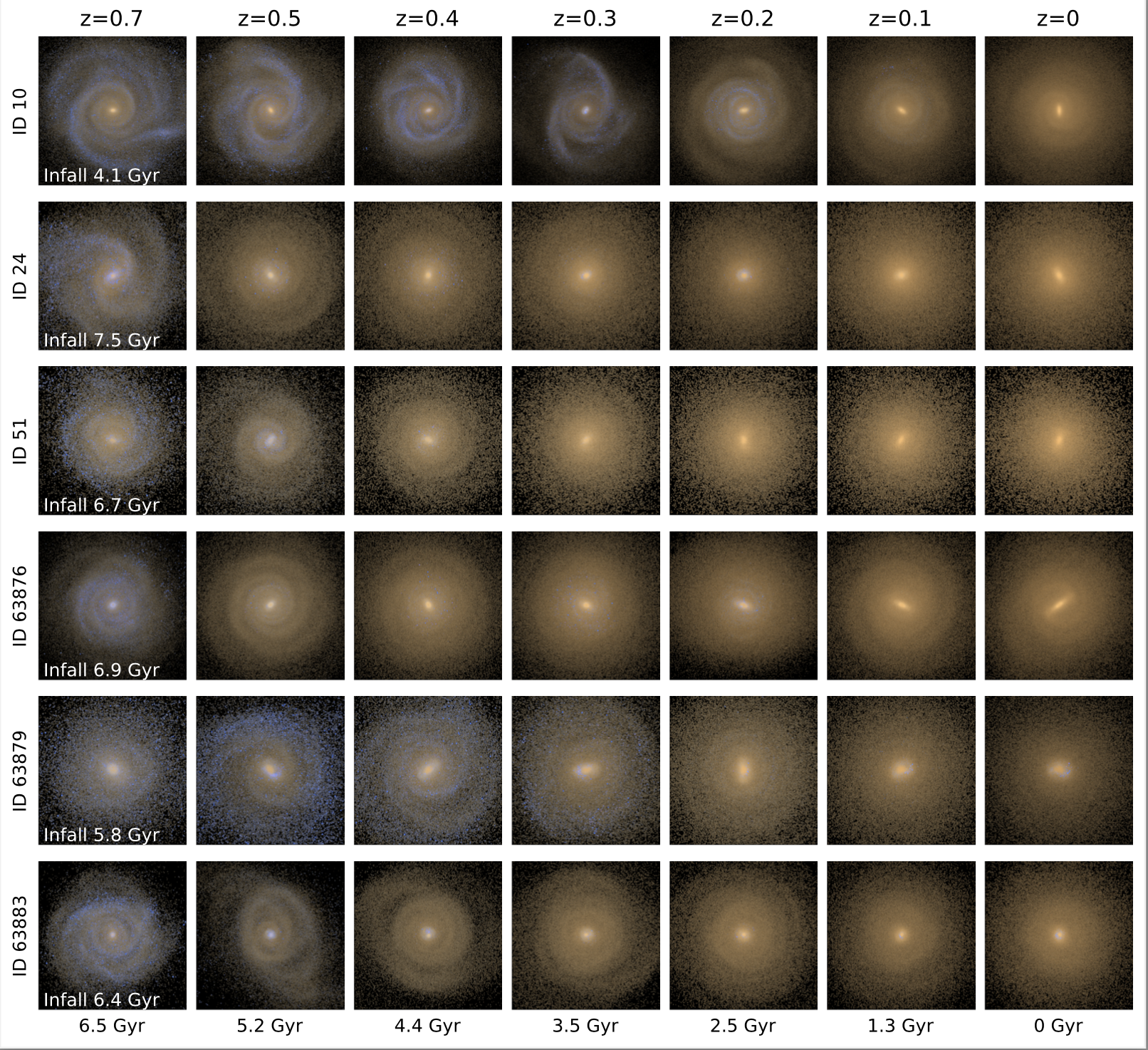}}
        \caption{Example of face-on SDSS (gri) synthetic images of the evolution of the subhalos from our 45 sample from TNG50. From left to right, the columns show the subhalos at $z=0.7,0.5,0.4,0.3,0.2,0.1$ and $z=0$. The ID of each subhalo is given (for z=0) as the y-label in each row. The left set of images correspond to examples of blue subhalos. The right set of images correspond to examples of our red subhalos sample (see Section \ref{subsec:BRG}).}
    \label{Fig2}
\end{figure*}

Motivated by these theoretical and observational considerations, in this work we use the TNG-50 cosmological hydrodynamical simulation \citep{nelson:18,pillepich:18,marinacci:18,naiman:18,springel:18} to investigate whether ram-pressure stripping alone, operating over extended timescales in low-mass group environments, can transform blue spiral galaxies into red spirals. We focus on the orbital and gas-depletion signatures of this slow-stripping process and assess whether the resulting systems resemble observed red spirals, potentially representing a transitional stage prior to S0 formation.

This article is organized as follows. In Section~\S\ref{sec:methods}, we describe the TNG50 simulation and our galaxy selection methodology. In Section~\S\ref{sec:results}, we present our results. Finally, Section~\S\ref{sec:conclusions} summarizes our conclusions.

%%%%%%%%%%%%%%%%%%%%%
\section{The TNG-50 simulation and the sample selection}
\label{sec:methods}
%%%%%%%%%%%%%%%%%%%%%
In the following sub-sections we briefly describe the TNG-50 simulation, and our sample galaxy-selection.

\subsection{Illustris TNG-50}
The Illustris TNG-50 is the simulation with the highest resolution in the  Illustris-TNG suite of cosmological-magneto-hydrodynamical simulations \citep{nelson:18,pillepich:18,marinacci:18,naiman:18,springel:18}. The TNG simulations are run with the Arepo MHD adaptive code \citep{springel:10}. The simulations use a a flat Lambda-CDM cosmological framework with matter density $\Omega_m=0.3089$, and Hubble constant $h=0.6774$, as given by the Planck data \citep{planck:16}.

The TNG simulations consist of a set of simulations with different domain sizes, and resolutions. In particular TNG-50 has a ($51.7$ Mpc)$^{3}$ volume box, with a baryonic mass resolution $m_b = 8.5\times10^4$ M$_{\odot}$, and a dark matter mass resolution $m_{DM} = 4.5\times10^5$ M$_{\odot}$. The minimum allowed adaptive gravitational softening length for gas cells (co-moving Plummer equivalent) is $\epsilon_{gas,min}= 74$ pc and for the stars and DM  $\epsilon_{DM,*}= 288$ pc.  

Given its relatively high mass resolution but limited volume, TNG50 is the optimal suite to study the interplay between low- and intermediate-mass galaxies and the field and group environments, whereas massive galaxy clusters are virtually absent from the simulated box.

The most important physical ingredients included in the TNG simulations are: The gas radiative processes, the star formation in the dense interstellar medium, the evolution of the stellar population and the chemical enrichment from supernovae Ia, II, as well as from AGB stars,  heating from a spatially homogeneous UV/X-ray background, galactic-scale stellar-driven outflows, and the seeding, growth and feedback of super massive black holes \citep{nelson:19}.

The star formation is modeled with a density threshold ($0.1$ cm$^{-3}$) as described in \cite{springel:03}. In such a star formation recipe the gas parcels are stochastically converted into star particles when their density is greater than $n_{H}=0.1$ cm$^{-3}$ \citep{kennicutt:83} on a time-scale proportional to the local dynamical time of the gas.

In summary, the TNG-50 simulation combines an intermediate-scale volume with a high mass resolution, providing an optimal tool for studying the transformation of blue spiral galaxies mainly by RPS.

It has no be noted that we use the term \textit{subhalo} to refer to a simulated galaxy in TNG-50 (which contains gas, stars, and dark matter). We use the term \textit{halo} to refer to the complete Group or Cluster of simulated galaxies in TNG-50. 

%%================================%%
%%            Figure 3            %%
%%         Cui et al. 24          %%
%%================================%%
\begin{figure}[t!]
    \centering
    {\includegraphics[width=0.45\textwidth]{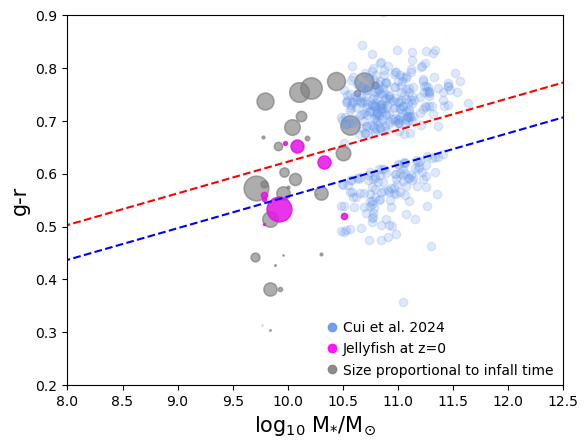}}
    \caption{The \textit{g-r} color as a function of stellar mass. The gray circles show our subhalo sample, where the size of the circles is proportional to the time since infall. The blue dots correspond to the massive red spiral galaxies reported in \cite{cui:24}. Following the classification given by \cite{cui:24}, the galaxies (and subhalos) which appear under the blue dashed line are classified as blue. Those which appear over the red dashed line are classified as red. And those in between the blue and red dashed lines are classified as in the green valley.  We plot in magenta those subhalos in our sample that at $z=0$ are classified as jellyfish in \cite{zinger:24}.}
\label{Fig5}
\end{figure}

%%================================%%
%%            Figure 4            %%
%%        KDE Infall time         %%
%%================================%%
\begin{figure}[t!]
    \centering
     {\includegraphics[width=0.4\textwidth]{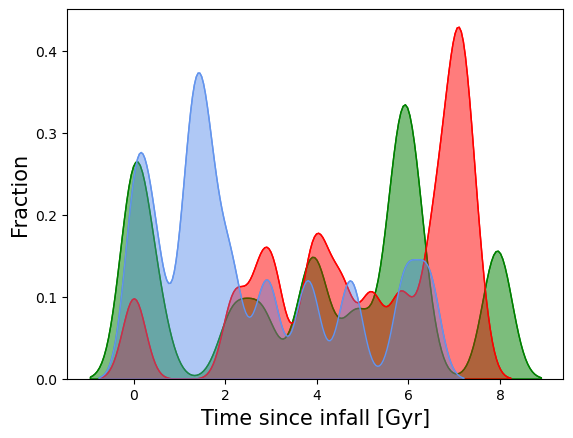}}
    \caption{Kernel density estimate of the infall times of subhalos in our sample. A Gaussian kernel is adopted, and the distributions are normalized to unit area. Red, green, and blue curves correspond to the red sequence, green valley, and blue subhalo populations (see Figure \ref{Fig5}), following the classification of \cite{cui:24}.}
    \label{Fig6}
\end{figure}

%%================================%%
%%          Figure 5              %%
%        Mass Assembly            %%
%%================================%%
\begin{figure*}[t!]
    \centering
    {\includegraphics[width=0.98\textwidth]{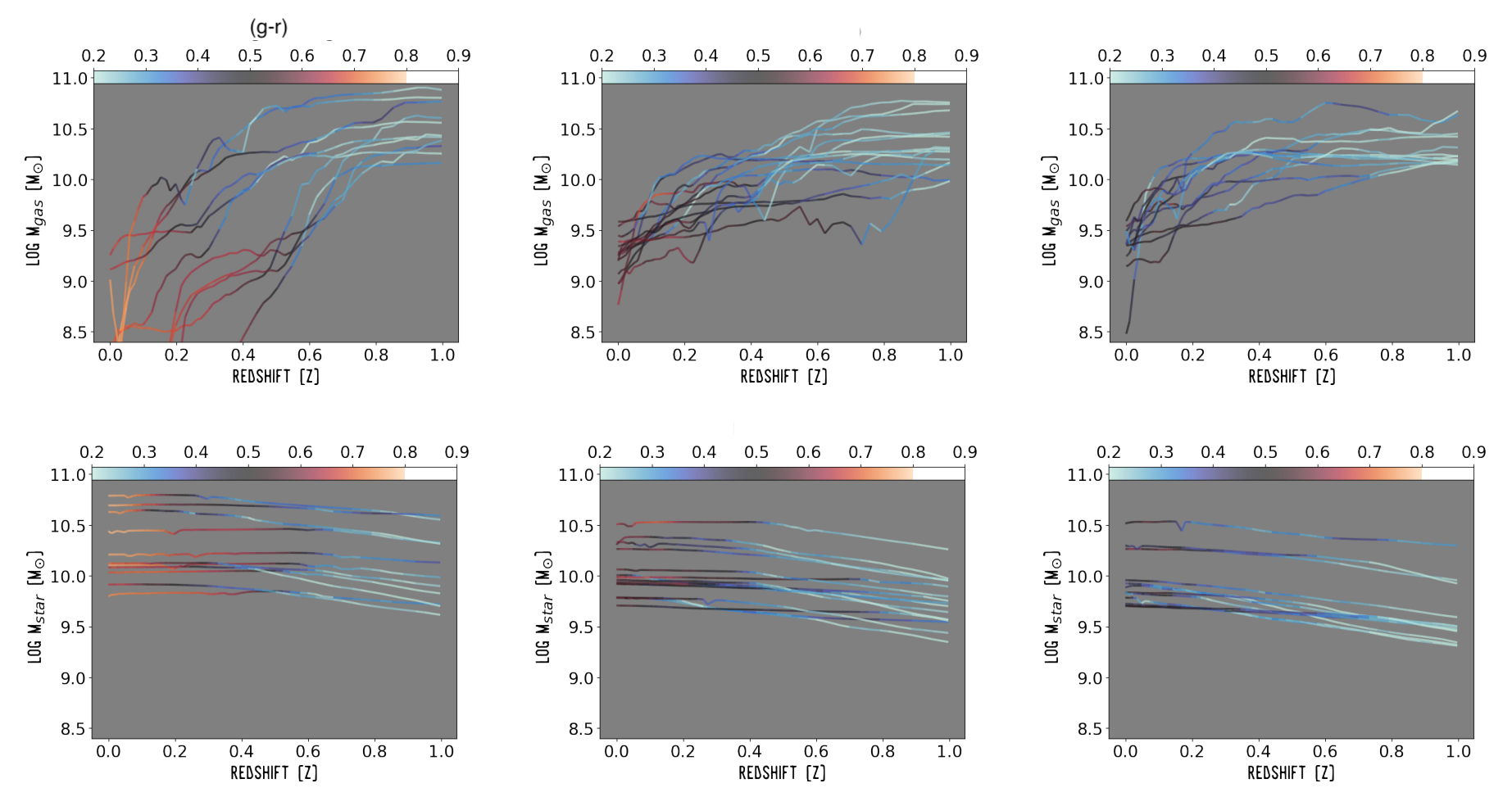}}
    \caption{Gas (top panels) and stellar (bottom panels) mass assembly for our subsample of gas-poor simulated disks from TNG50. The left, central and right columns show systems classified at z=0 as blue, green, and red, respectively.}
    \label{Fig3}
\end{figure*}

%%================================%%
%%          Figure 6              %%
%         (g-r) & SFR             %%
%%================================%%
\begin{figure*}[t!]
    \centering
     {\includegraphics[width=0.32\textwidth]{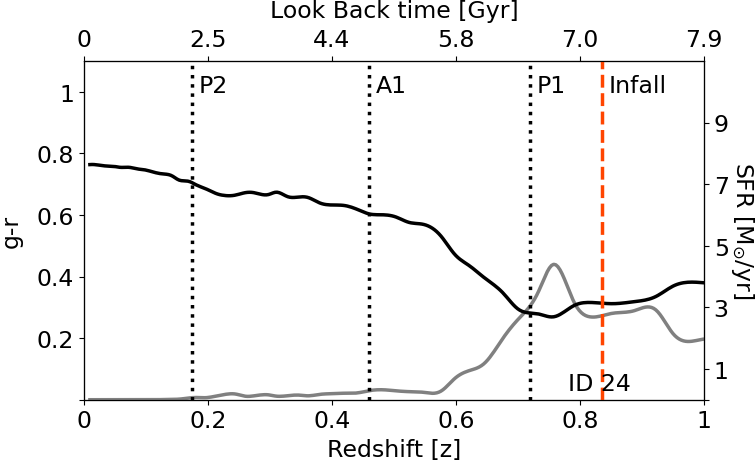}}
     {\includegraphics[width=0.32\textwidth]{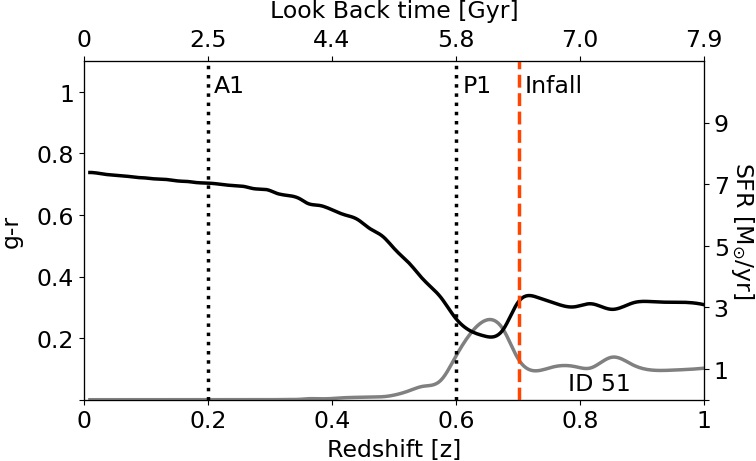}}
     {\includegraphics[width=0.32\textwidth]{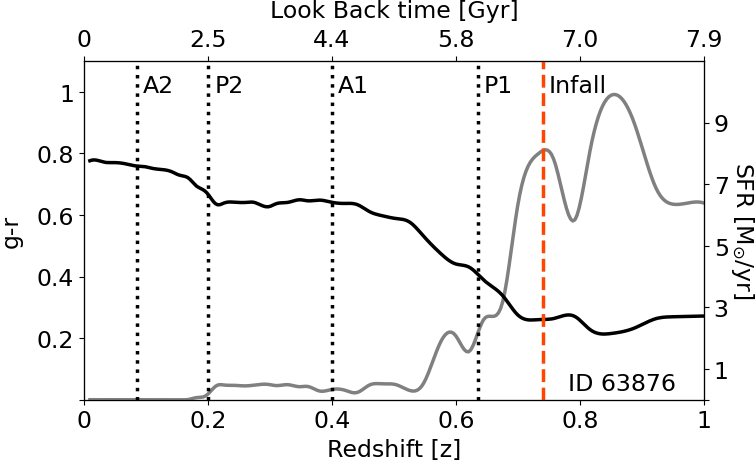}}\\
     {\includegraphics[width=0.32\textwidth]{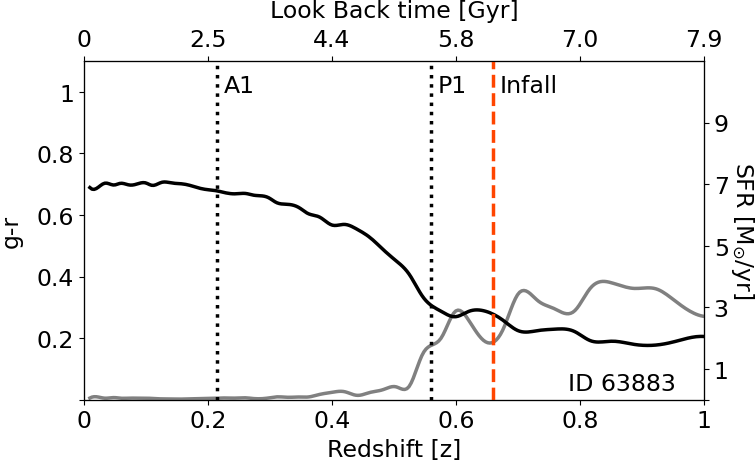}}
     {\includegraphics[width=0.32\textwidth]{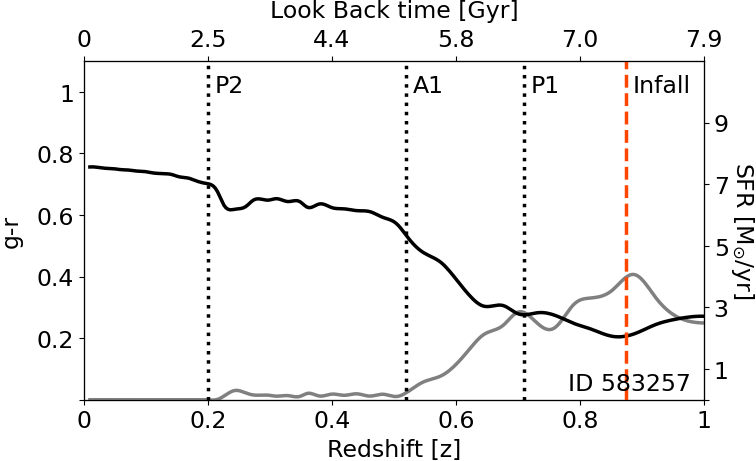}}
     {\includegraphics[width=0.32\textwidth]{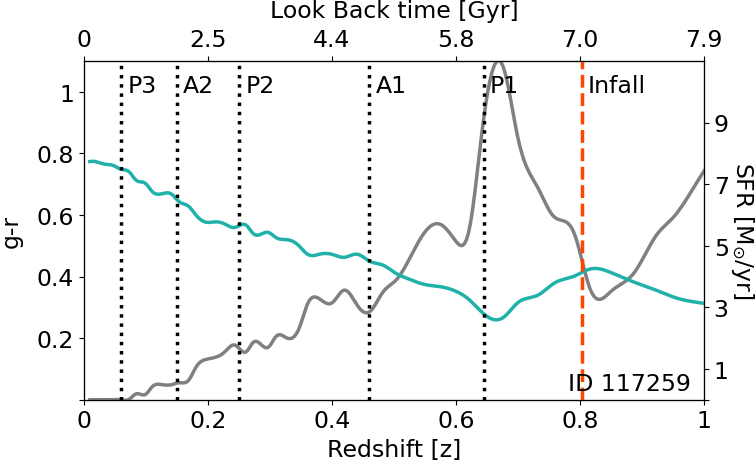}}\\
     {\includegraphics[width=0.32\textwidth]{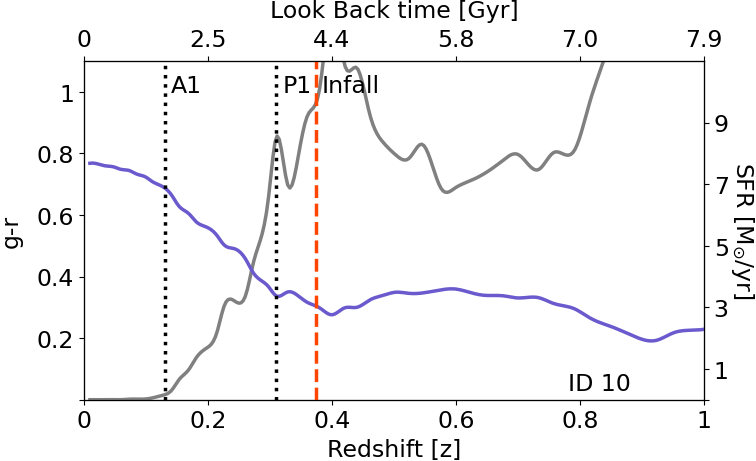}}
     {\includegraphics[width=0.32\textwidth]{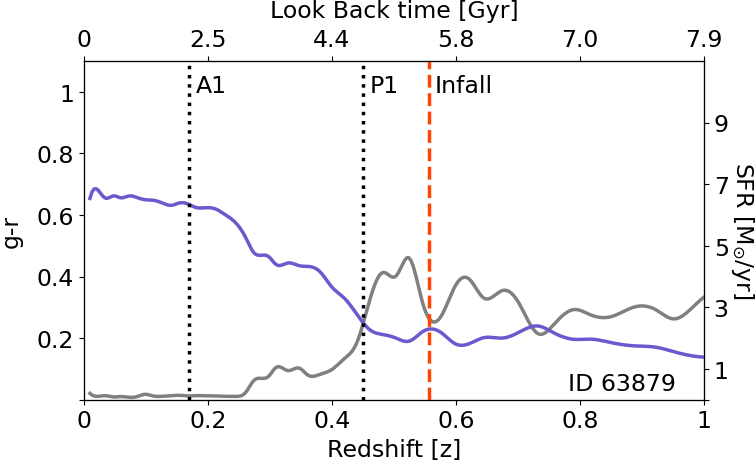}}
     {\includegraphics[width=0.32\textwidth]{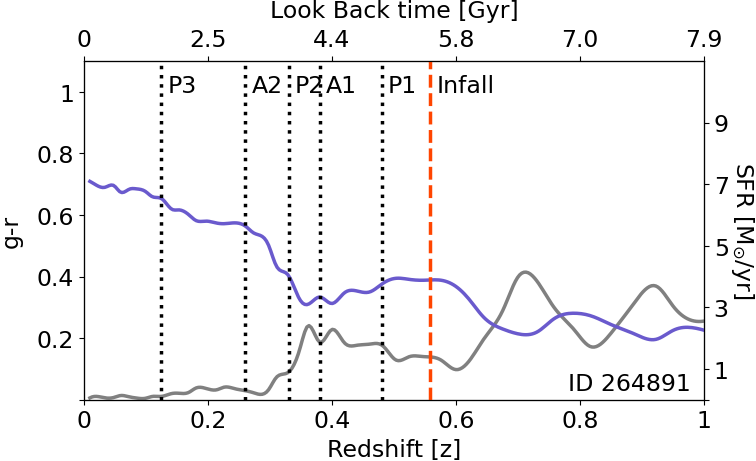}}\\
     {\includegraphics[width=0.32\textwidth]{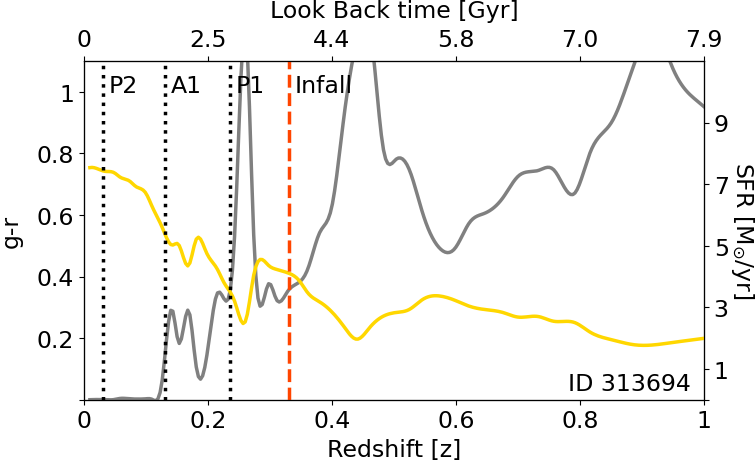}}
     {\includegraphics[width=0.32\textwidth]{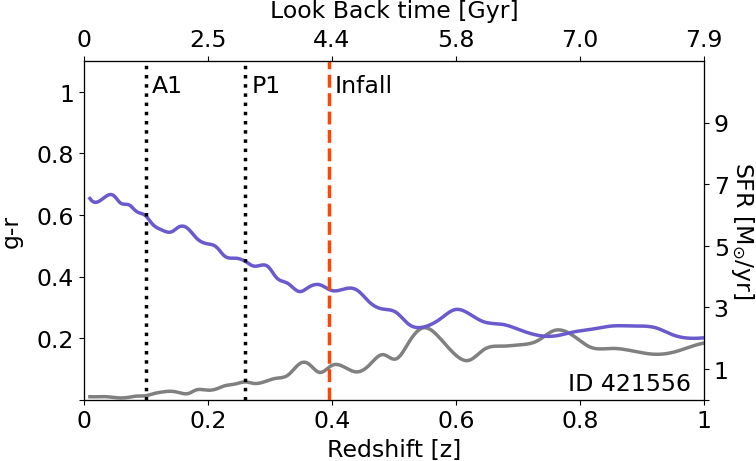}}
    \caption{Color evolution and star formation history for our subsample of 11 red disk galaxies. The SFR is traced by the light gray curve and indicated on the right y-axis. The g-r color is traced by black, blue, cyan, or yellow curves. The red vertical line shows the infall time. The vertical dashed black lines show the peri- and apo-clustercentric distances.}
    \label{Fig7}
\end{figure*}

%%%%%%%%%%%%%%%%%%%%%%%%%%%%%%%%
\subsection{Sample selection}
\label{subsecsec:sample}
%%%%%%%%%%%%%%%%%%%%%%%%%%%%%%%%
% REFEREE COMMENT 3 (re arrange)
\textbf{S0 and red/anemic spirals exhibit disk-dominated morphologies, old stellar populations, and low gas fractions \citep{masters:10}. To construct our simulated sample at $z=0$, we selected galaxies with well-defined stellar disks, without initially imposing a central/satellite restriction. We use the kinematic morphology catalog of \cite{zana:22}, which provides stellar kinematic decompositions for subhalos with $M_* > 10^9 M_{\odot}$. We select subhalos in which at least 50\% of the total stellar mass resides in the thin disk component ($M{_{thin}}/M_{} > 0.5$). This criterion yields 963 subhalos at $z=0$.}

% REFEREE CPOMMENT 2
\textbf{In order to obtain a sample of robustly resolved, gas-poor disk galaxies, we further select subhalos with stellar masses larger than $5\times10^9\,M_{\odot}$. Although TNG50 resolves galaxies below this mass, this conservative threshold avoids the lower-mass regime where the stellar component contains fewer resolution elements and where structural, kinematic, and star formation measurements become increasingly affected by stochastic variations. The selected galaxies are additionally required to have gaseous masses below $5\times10^9\,M_{\odot}$, ensuring that their baryonic content of the selected systems is dominated by stars. These criteria reduce the sample to 45 subhalos, but provide a population of well defined disk galaxies for which the evolution of the stellar component and gas content can be reliably followed over several Gyr. We note that they all classify as satellite galaxies}.

%Mergers and Flybys
To isolate transformations driven by ram pressure stripping alone, we quantify the stellar mass acquired through mergers and flybys between $z=1$ and $z=0$ (a lookback time of $\sim8$ Gyr). Merger contributions are obtained directly from the SubLink merger trees by summing the stellar mass accreted through identified merger events. The flybys contribution are those stellar particles that were previously associated with a distinct subhalo, but later become gravitationally bound to the galaxy without a formal merger event in the merger tree. This procedure captures stellar mass exchange during non-merging close encounters.

The top panel of Figure~\ref{Fig1} shows, for each of the 45 subhalos, the stellar mass fraction accreted via mergers (blue points) and flybys (magenta points), normalized to the total stellar mass at $z=0$. Gray horizontal lines indicate 5\%, 1\%, and 0.1\% mass fractions.
Only four subhalos acquire more than 5\% of their final stellar mass through mergers since $z=1$. We therefore restrict our analysis to galaxies with $<5\%$ stellar mass growth from mergers and flybys combined, removing these four objects and leaving a final sample of 41 subhalos.

% Justification of the 5% Threshold
A stellar mass growth below 5\% corresponds to minor mergers (interactions) with mass ratios $\lesssim 1{:}20$. Analytic arguments and controlled numerical simulations have demonstrated that disk heating and bulge growth scale approximately with the satellite-to-host mass ratio, such that mergers in this regime induce only modest thickening and limited morphological transformation (e.g., \citealt{toth:92}; \citealt{quinn:93}; \citealt{hopkins:09}; \citealt{purcell:10}). In particular, mergers with mass ratios $\lesssim 1{:}10$ are generally inefficient at transforming rotationally supported disks into dispersion-dominated systems, and the structural response of the disk scales with the fractional accreted mass and associated orbital energy deposition (\citealt{hopkins:09}). Therefore, stellar mass growth below 5\% is not expected to produce substantial global morphological change.
Although extreme orbital configurations can enhance local perturbations, encounters with mass ratios $\lesssim 1{:}20$ lack sufficient dynamical impact to significantly disrupt a pre-existing thin disk. Nonetheless, to verify that gravitational interactions do not alter the dynamical structure of our galaxies, we directly measure their rotational support over time.

For each system, we compute the stellar rotational support parameter $\kappa_{\mathrm{rot}}$ at every snapshot between $z=1$ and $z=0$. $\kappa_{\mathrm{rot}}$ is defined as the fraction of the total stellar kinetic energy invested in ordered rotation \citep{sales:12},
\begin{equation}
    \kappa_{rot} = \frac{K_{rot}}{K} = \frac{\Sigma_i \frac{1}{2} m_i (v_{\phi,i})^2}  {\Sigma_i \frac{1}{2} m_i v_i^2} \mbox{ ,}
\end{equation}
where $m_i$ is the mass of stellar particle $i$, $v_i$ its total velocity relative to the galaxy center, and $v_{\phi,i}$ its azimuthal velocity component measured with respect to the stellar angular momentum axis. Systems with $\kappa_{\mathrm{rot}} \gtrsim 0.5$ are typically classified as rotationally supported disks.

%We find no systematic decline in $\kappa_{\mathrm{rot}}$ across our sample from $z=1$ to $z=0$, indicating that the stellar components retain their dynamical structure over the last $\sim8$ Gyr. Therefore, we conclude that gravitational interactions have a negligible contribution to the structural evolution of these galaxies during this period, allowing us to isolate environmentally driven gas-removal processes as the primary quenching mechanism. While RPS is likely the dominant effect, (as previously mentioned) other mechanisms such as viscous stripping, thermal evaporation, and starvation may also contribute, acting together to deplete the gas reservoirs of these systems.

% REFEREE COMMENT 4 (IIDES)
% modify paragraph 
\textbf{We find no systematic decline in $\kappa_{\rm rot}$ across our sample from $z=1$ to $z=0$, indicating that the stellar components retain their dynamical structure over the last $\sim8$ Gyr. Together with the negligible stellar mass growth through mergers and flybys, this suggests that strong gravitational perturbations have not played a dominant role in the evolution of these galaxies. We therefore interpret the observed gas depletion as being primarily driven by hydrodynamical environmental processes. Nevertheless, we cannot completely exclude the influence of other environmental mechanisms, such as the global tidal field of the host halo, starvation, viscous stripping, or thermal evaporation, which may contribute to weakening the gaseous component and act together with RPS. Our selection criteria are designed to minimize the effects of gravitational interactions, allowing us to isolate systems in which RPS is expected to dominate the evolution of the cold gas reservoir.}

%Host Halo Masses
% REFEREE COMMENT 1
\textbf{The bottom panel of Figure~\ref{Fig1} shows the host halo mass as a function of the stellar mass for the 41 galaxies in our sample. Owing to its relatively small cosmological volume ($\sim50~{\rm Mpc})^3$, TNG50 predominantly samples field galaxies and group environments, with only a limited number of low-mass clusters. The galaxies in our final sample span a broad range of environments, from massive galaxy halos through galaxy groups to low-mass clusters ($M_{\rm halo}\sim10^{11.8}$--$10^{14.3}\,M_{\odot}$), although the majority reside in group-scale halos. This environmental diversity allows us to investigate the transformation of disk galaxies across the regime where RPS is expected to transition from relatively weak to increasingly efficient.}

%%%%%%%%%%%%%%%%%%%%%%%%%%%%%%%%%%
\section{Results}
\label{sec:results}
%%%%%%%%%%%%%%%%%%%%%%%%%%%%%%%%%%%
\subsection{Sample Selection and Morphology}
We selected galaxies with low gas mass that retain a disky morphology at $z=0$, enabling a clean study of the transformation from blue to red spirals associated with ram-pressure-driven gas removal.
Figure~\ref{Fig2} shows synthetic SDSS (gri) face-on images for representative subhalos at multiple snapshots. By $z=0$, all galaxies appear redder, while preserving disk morphology (see also Figure~\ref{Fig5}). We applied the morphological classification, following \citet{marasco:23} to confirm that the galaxies in the final sample are spirals. The morphological classification is based on stellar mass, B-V color, and structural indicators (concentration and clumpiness) computed from synthetic V-band images and calibrated against the MORPHOT observations \citep{fasano:12,vulcani:23}. All galaxies in the final sample retain a spiral morphology at $z=0$.\\

%--------------------------------------------%
\subsection{Color Evolution: Red, Green, and Blue Spirals}
\label{subsec:BRG}
%--------------------------------------------%
The \textit{g-r} colors of the subhalos were computed from $z=1$ to $z=0$. Figure~\ref{Fig5} shows the color–stellar mass distribution (gray circles), with blue- and red-spiral regions indicated following \citet{cui:24} (blue and red dashed lines) and the green valley between them. Observational points from \citet{cui:24} are over-plotted for comparison; overlap occurs only for the most massive subhalos.  

Galaxies below the blue dashed line in the color-mass diagram (Figure ~\ref{Fig5}) are classified as \textit{blue}, above the red dashed line as \textit{red}, and intermediate as \textit{green}. Figure~\ref{Fig6} shows the infall time distributions for each color class. In our sample, 30\% of the galaxies are blue, 35\% green, and 35\% red. About 60\% of blue subhalos have infall times $<2$~Gyr, consistent with their ongoing star formation. Conversely, 93\% of red subhalos have infall times $>2$~Gyr, with 75\% exceeding 3.5~Gyr, 50\% exceeding 6~Gyr, and 25\% exceeding 7~Gyr. While morphological transformation via RPS is rapid in high-density environments (1–2~Gyr; \citealt{deeley:21}), our results demonstrate that in low-mass clusters and groups, red spirals can form over much longer timescales, up to 7~Gyr after infall.

%--------------------------------------------%
\subsection{Gas and Stellar Mass Assembly}
%--------------------------------------------%
Figure~\ref{Fig3} presents the gas (top row) and stellar (bottom row) mass assembly for red, green, and blue subhalos. Gas mass decreases steadily after $z=1$, reaching $M_{\rm gas} \lesssim 10^{8.5}$ M$_\odot$ for red spirals and $M_{\rm gas} \lesssim 3\times10^9$ M$_\odot$ for green and blue spirals. Stellar mass increases slowly until $z\sim0.4$ and then stabilizes, while colors evolve toward redder values. These trends indicate that hydrodynamical gas removal, primarily through RPS, dominates the quenching
in these systems.

To verify that gas loss is primarily due to RPS rather than star formation (starvation), we compared the gas removal rate with star formation consumption. Gas is lost in impulsive episodes correlated with pericentric passages, exceeding star-formation-driven depletion (see Figure~\ref{Fig:appendix}). Hence, RPS dominates the quenching of the sample. From this point onward, we focus on red subhalos, which have completed their transformation.

%%%%%%%%%%%%%%%%%%%
%%    Figure 7   %%
%%  Diagram F_s  %%
%%%%%%%%%%%%%%%%%%%
\begin{figure}[t!]
    \centering
     {\includegraphics[width=0.28\textwidth]{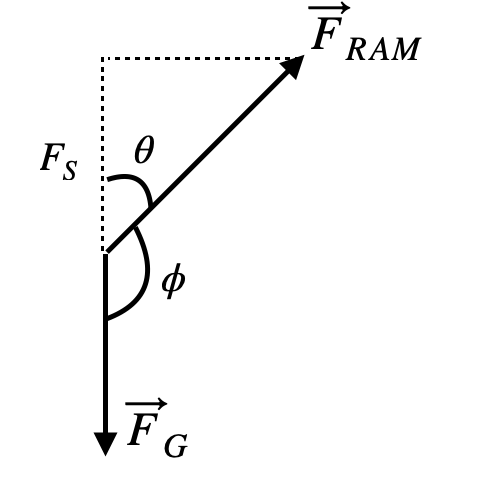}}
     {\includegraphics[width=0.18\textwidth]{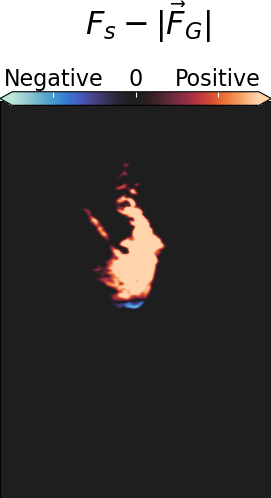}}
    \caption{Left: Diagram to define $F_{S}$; the component of the RAM force ($\vec{F}_{RAM}$) projected over the direction of the gravitational force ($\vec{F}_{G}$). Right: Example of $F_S - |\vec{F}_{G}|$ map in a subhalo.}
    \label{Fig:diagram}
\end{figure}

%%%%%%%%%%%%%%%%%%%
%%    Figure 8   %%
%%  Diagram F_s  %%
%%%%%%%%%%%%%%%%%%%
\begin{figure*}[t!]
    \centering
     {\includegraphics[width=0.2\textwidth]{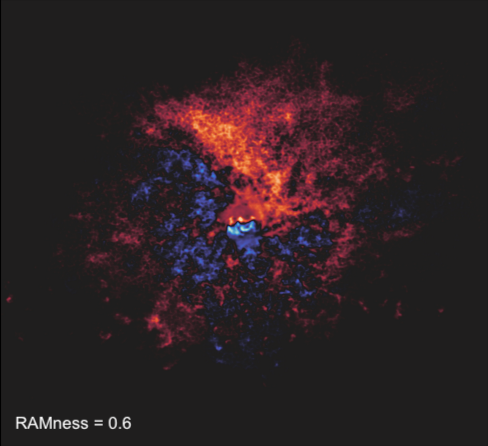}}
     {\includegraphics[width=0.2\textwidth]{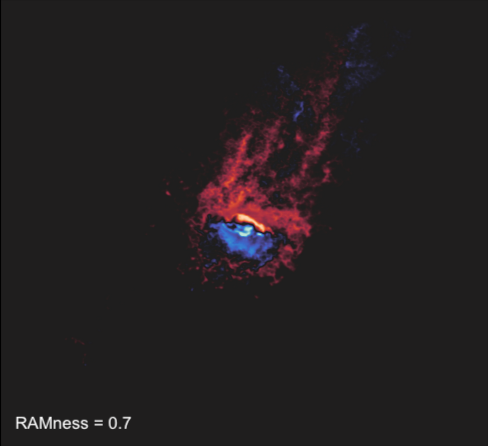}}
     {\includegraphics[width=0.2\textwidth]{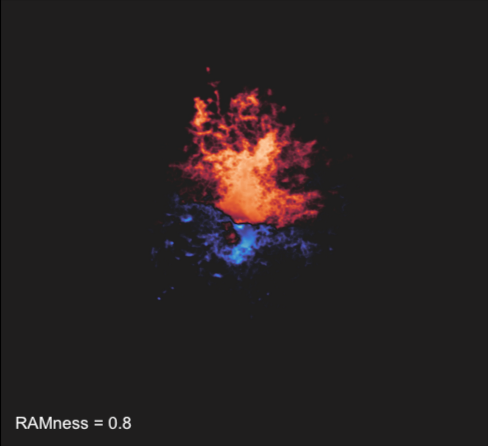}}
     {\includegraphics[width=0.2\textwidth]{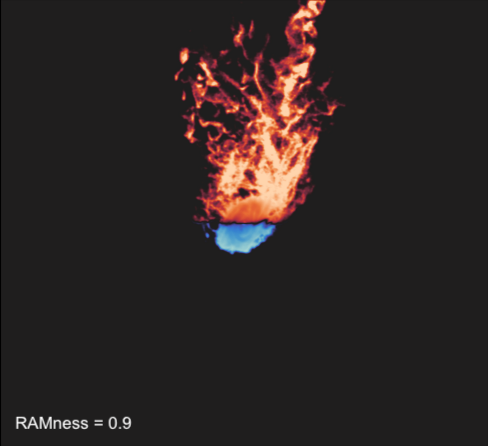}}
    \caption{Random snapshots of subhalos with increasing RAMness parameter: $F_S - |\vec{F}_{G}|$, color coded as in Figure \ref{Fig:diagram}.\\}
    \label{Fig:ramness_evol}
\end{figure*}

%%%%%%%%%%%%%%%%%%%%%%%%%%
%%      Figure 9        %%
%% RAMness Reds > 6Gyr  %%
%%%%%%%%%%%%%%%%%%%%%%%%%%
\begin{figure*}[t!]
    \centering
     {\includegraphics[width=0.48\textwidth]{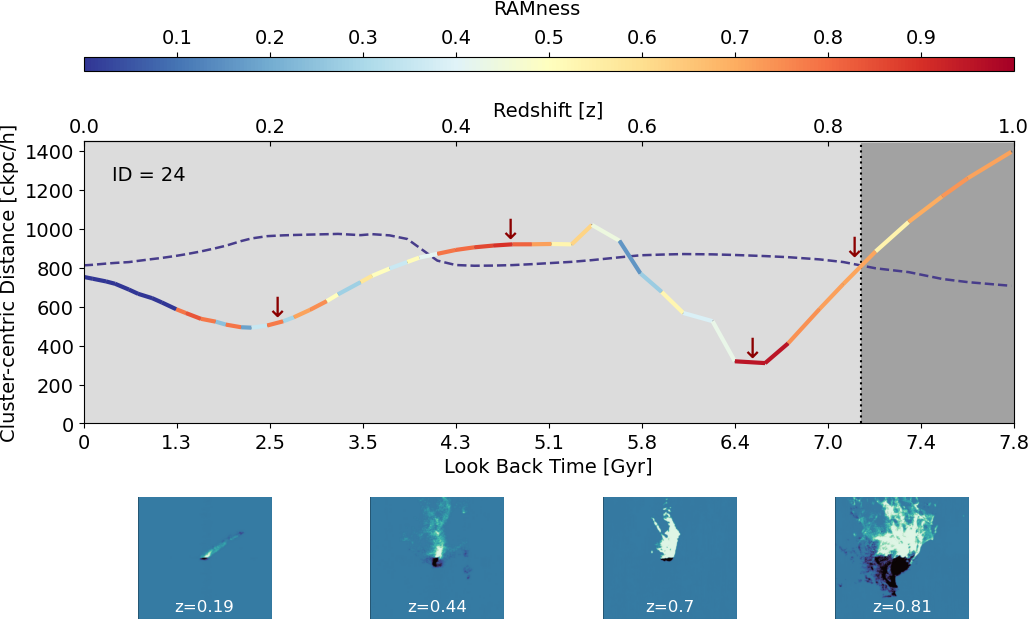}}
     {\includegraphics[width=0.48\textwidth]{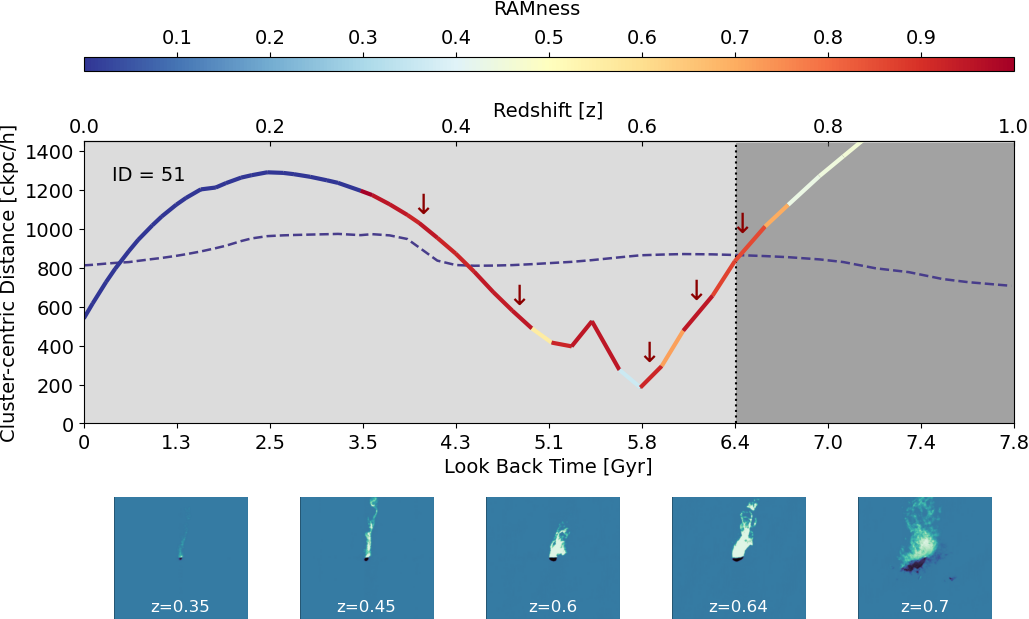}}\\
     {\includegraphics[width=0.48\textwidth]{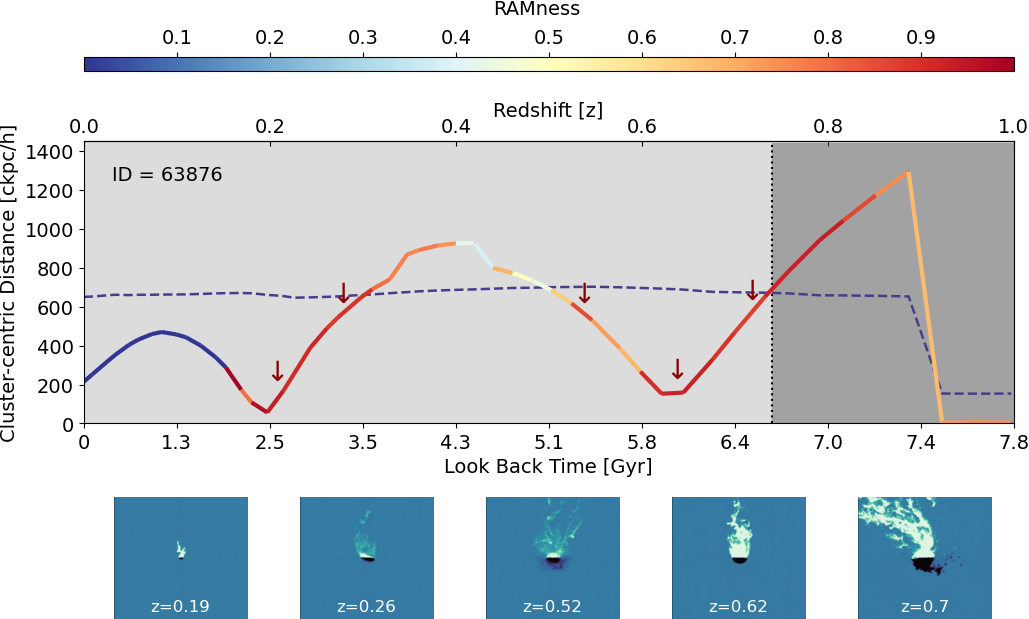}}
     {\includegraphics[width=0.48\textwidth]{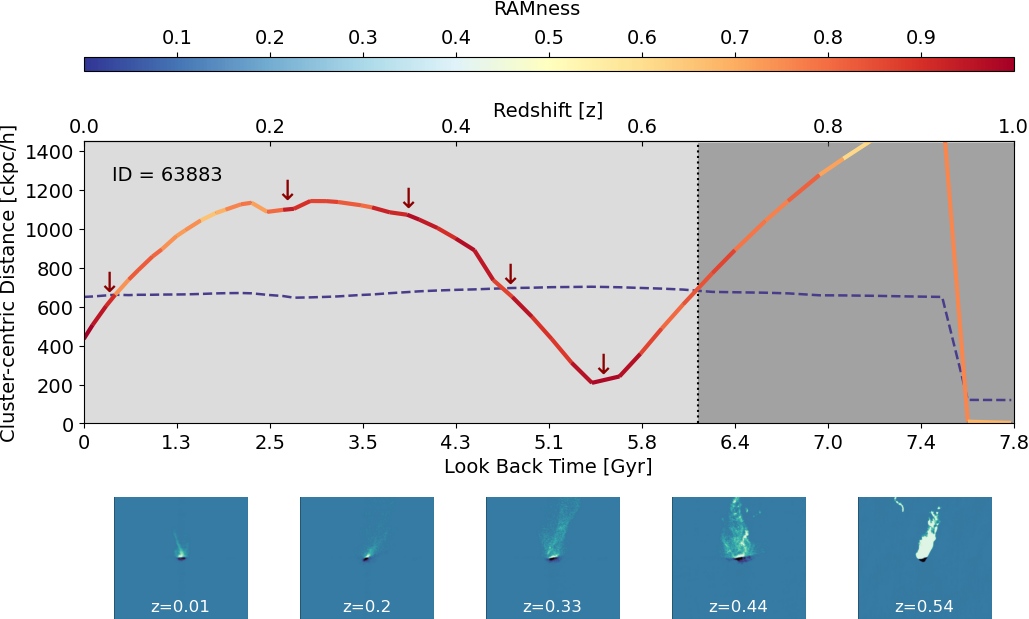}}\\
     {\includegraphics[width=0.48\textwidth]{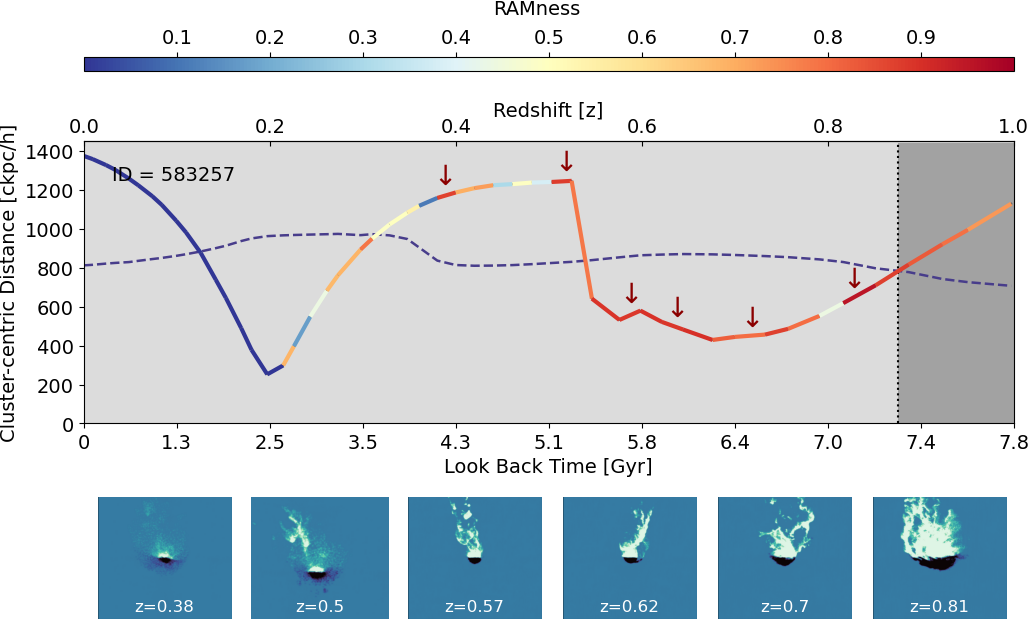}}   
     {\includegraphics[width=0.48\textwidth]{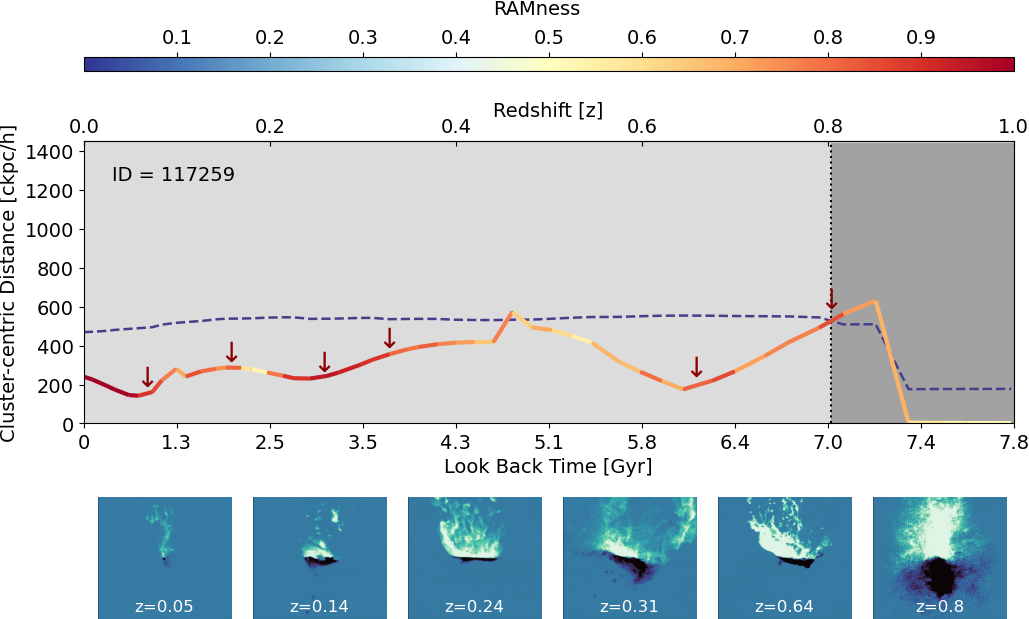}}
    \caption{Cluster-centric distance as a function of lookback time of those red subhalos in our sample which have an infall time $\ge 6$ Gyr. It is color-coded by the RAMness parameter defined in Equation 5.  The dashed curve shows the virial radius R200 of the group. The vertical dotted line marks the the infall time of the subhalo. The bottom panel shows some snapshots of the gas distribution at a specific time (marked by downward arrows in the main panel) where tentacles of gas are clearly seen.}     
    \label{Fig:Ramness0}
\end{figure*}
\vspace{0.3cm}

%----------------------------------------------------------%
\subsection{Slow Transformation from Blue to Red Spirals}
%----------------------------------------------------------%
\label{subsec:red_transformation}
We restrict our analysis to the 11 subhalos classified as red at $z=0$ (see Figure \ref{Fig5}). Figure \ref{Fig7} shows their evolution of \textit{g-r} color (left y-axis) and SFR (right y-axis) as a function of redshift and look back time. The vertical red dashed line indicate the infall time, while the black dashed lines mark peri- and apo-cluster passages.
Our red subhalos exhibit two distinct behaviors. Those represented with black solid lines in Figure \ref{Fig7} have spent the longest time within their host halo ($\gtrsim6$ Gyr since infall). After infall, these subhalos approach their first peri-cluster passage, showing an oscillating \textit{g-r} color of $\sim0.2-0.3$ mag. As they move from the first peri- to the first apo-center, \textit{g-r} rapidly increases to $\sim0.6$mag over $\sim1$ Gyr. Subsequently, as the subhalos continue their orbit toward the second peri-center passage, \textit{g-r} increases gradually and continuously, independent of orbital phase, eventually reaching $\sim0.7-0.8$ mag, completing their transformation into red spirals over $>4$ Gyr.
The SFR of these black-marked subhalos peaks at $\sim3$ M$_{\odot}/$yr before infall. After the first peri-center passage ($\sim5$ Gyr look back time), SFR briefly increases, then drops sharply as the subhalos move toward the first apo-center and remains low thereafter.
A notable case is the turquoise subhalo (ID 117259), which was accreted onto its group at $\sim7$ Gyr look back time. Near its first peri-center, it experiences a brief rejuvenation: \textit{g-r} decreases for $\sim1$ Gyr while the SFR rises from $\sim4$ to $\sim11$ M$_{\odot}/$yr. 
Afterwards, \textit{g-r} steadily increases at a roughly constant rate, taking $\sim6$ Gyr to reach $\sim0.8$ mag at $z=0$, demonstrating a slow, long-term transformation.
% REFEREE about turquoise subhalo
\textbf{We note that the temporary enhancement of the SFR observed in subhalo 117259 after the first pericentric passage is associated with ram-pressure induced compression of its initially massive and extended gas disk. Subsequent passages occur after substantial gas depletion and disk contraction, preventing similar rejuvenation episodes.}

Red subhalos with shorter infall times ($<6$ Gyr; purple lines in Figure \ref{Fig7}) show a similar but faster evolution. Their \textit{g-r} increases after the first peri-center passage reaching $\sim0.65-0.7$ by $z=0$. The SFR oscillates while declining gradually throughout this period.
A second special case, the yellow subhalo (ID 313694), was accreted onto its group at $\sim3.7$ Gyr look back time. Interestingly, its \textit{g-r} increases even before infall ($z\sim0.45$–$0.3$), then briefly decreases as it nears the first peri-center passage, corresponding to a SFR spike from $\sim3$ to $\sim12$ M$_{\odot}/$yr. Subsequently, \textit{g-r} rapidly increases as the subhalo approaches its first apo-center.
Overall, these cases illustrate that the transformation from blue to red spirals in low-mass clusters and group environments is gradual, often spanning several Gyr, and can be temporarily modulated by bursts of star formation triggered during orbital passages.

% Referee COMMENT 2
\textbf{We further investigate whether the transformation timescale depends on host halo mass. The 11 red spiral galaxies span a wide range of environments, residing in halos with masses from $6.7\times10^{11}$ to $2.2\times10^{14}\,M_{\odot}$, thus covering both group-scale halos and low-mass cluster regimes. To test for a possible dependence, we compute the Spearman rank correlation between transformation time and host halo mass and find $\rho = 0.07$ with $p = 0.83$, indicating no statistically significant monotonic correlation. This result indicates that the transformation timescale is not primarily set by host halo mass, but is instead more closely linked to the orbital histories of the galaxies and their cumulative exposure to RPS along their orbits.}

% REFEREE COMMENT 2 Wetzel+ 13
\textbf{Interestingly, our lack of a significant correlation between transformation time and halo mass is broadly consistent with the observational results of \citeauthor{wetzel:13} \citeyear{wetzel:13}, suggesting that orbital history may be more important than host halo mass within the range explored here.}

%-------------------------------------% 
\subsection{RAMness in the Red sample} 
\label{sec:ramness} 
%-------------------------------------%
As described previously, our red subhalo sample was carefully selected to exclude any gravitational interactions (such as mergers or flybys) that could influence the transformation from blue to red spirals. Therefore, within the constraints of our selection criteria, the observed evolution of these galaxies can be attributed mainly to RPS.

%%%%%%%%%%%%%%%%%%%%%%%%%%
%%      Figure 10       %%
%% RAMness Reds < 6Gyr  %%
%%%%%%%%%%%%%%%%%%%%%%%%%%
\begin{figure*}[t!]
    \centering
     {\includegraphics[width=0.48\textwidth]{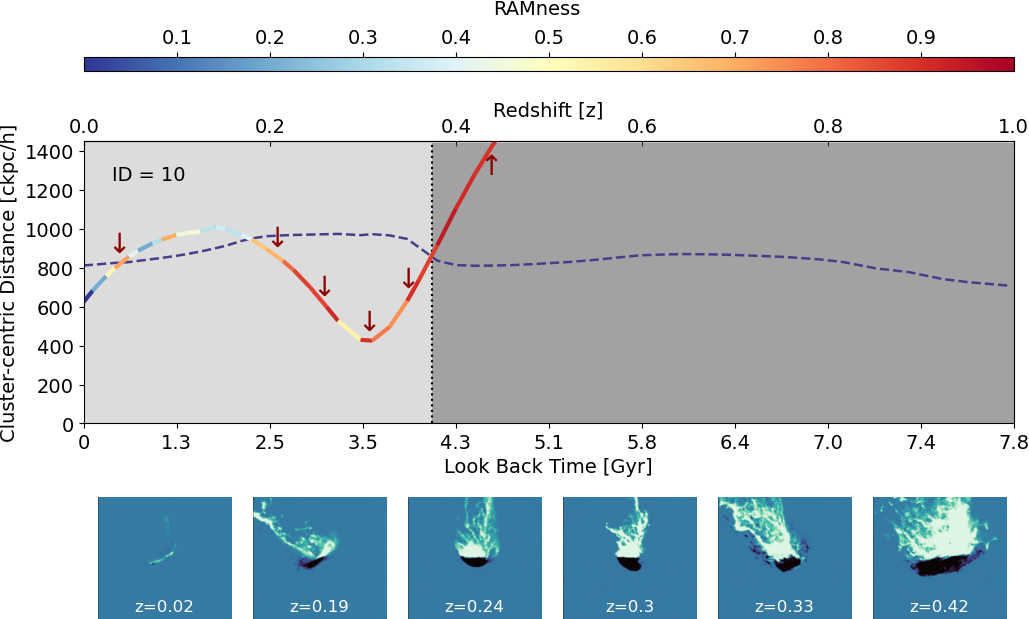}}
     {\includegraphics[width=0.48\textwidth]{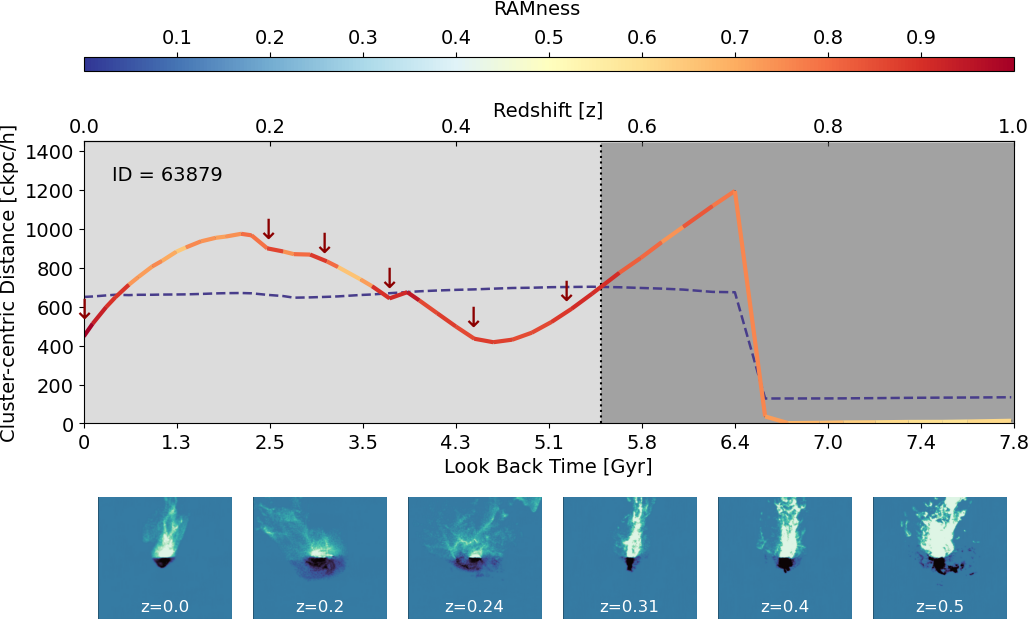}}\\
     {\includegraphics[width=0.48\textwidth]{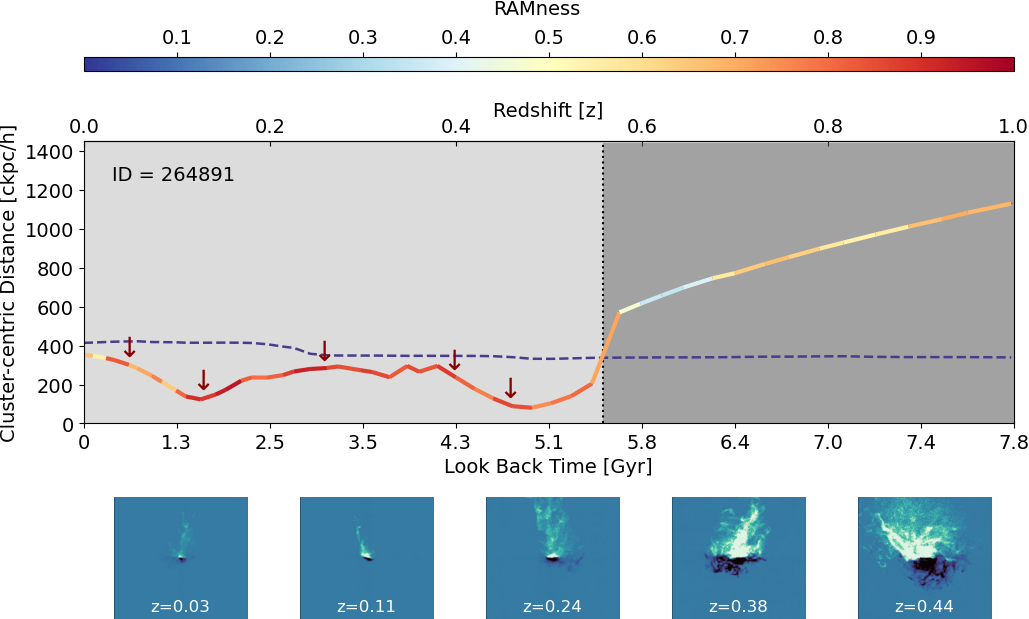}}
     {\includegraphics[width=0.48\textwidth]{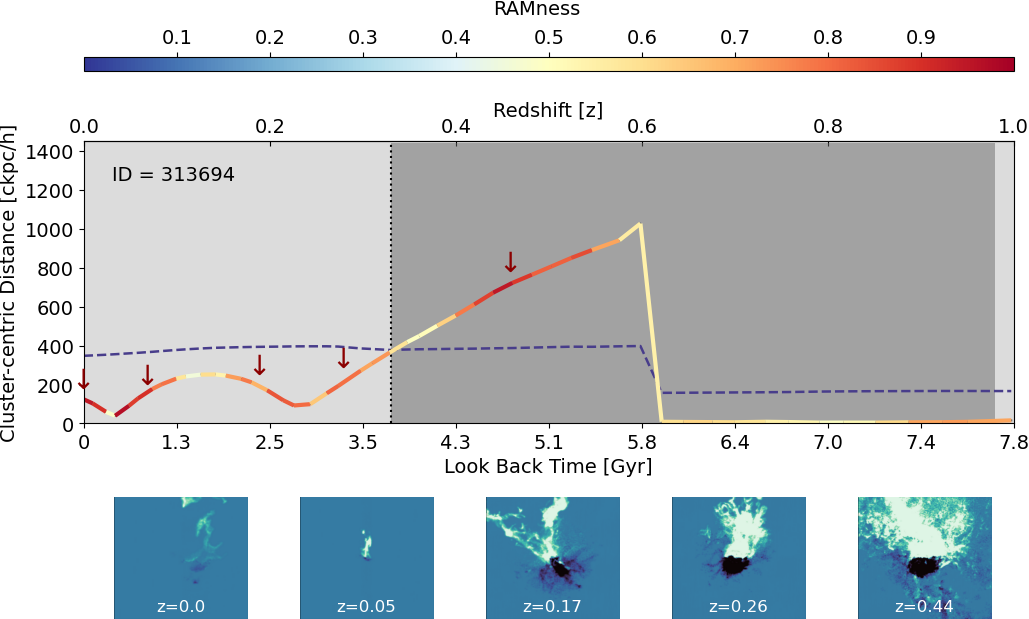}}\\
     {\includegraphics[width=0.48\textwidth]{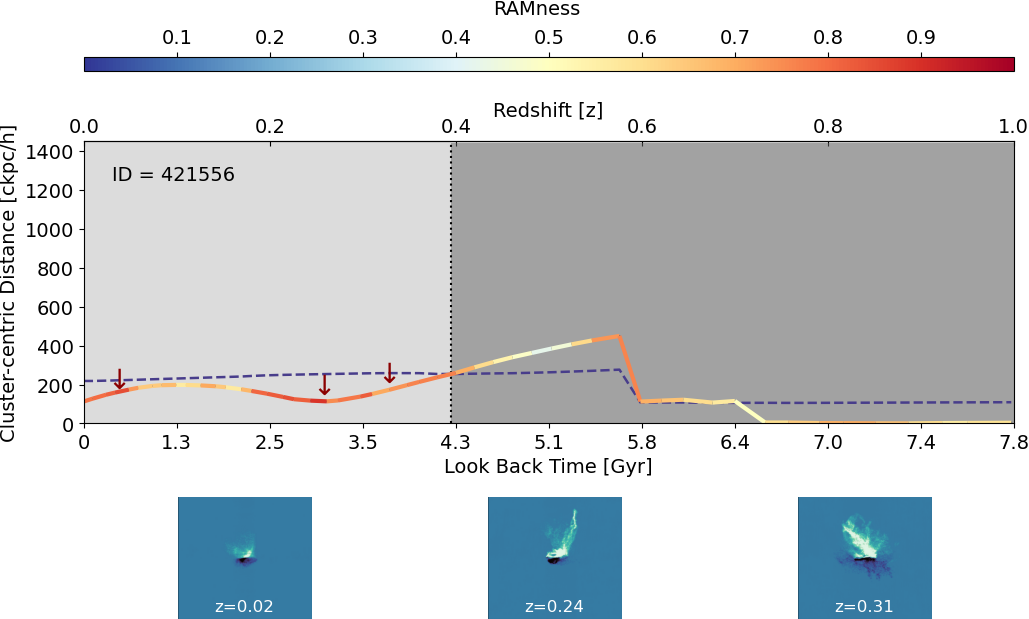}}
    \caption{Same as Figure \ref{Fig:Ramness0} but for subhalos with infall times $< 6$ Gyr.}
    \label{Fig:Ramness1}
\end{figure*}

The ram pressure experienced by a subhalo is defined following \cite{gunn:72} as:
\begin{equation}
P_{\rm RAM} = \rho_{\rm ICM} v_{\rm rel}^2 \mbox{ ,}
\end{equation}
where $v_{\rm rel}$ is the relative velocity between the subhalo gas and the ICM.

For each subhalo, we compute the local ICM properties within a sphere of radius 150 kpc centered on the subhalo \citep[see][]{yun:19}. This yields the mean ICM density $\rho_{\rm ICM}$ and the bulk velocity $\vec{v}_{\rm ICM}$, which are then used to calculate the ram pressure on each gas particle $j$, $P_{\rm RAM, j}$.

The corresponding ram force on each gas particle is computed assuming a circular cross-section with radius equal to the softening length of the gaseous component in TNG-50 ($A_j$):
\begin{equation}
\vec{F}_{\rm RAM,j} = P_{\rm RAM,j} A_j \frac{\vec{v}_{\rm ICM}-\vec{v}_j}{|\vec{v}_{\rm ICM}-\vec{v}_j|} \mbox{ ,}
\end{equation}
where $\vec{v}_j$ is the particle $j$ velocity. The gravitational force exerted on the particle $j$ by the subhalo, $\vec{F}_{\rm G,j}$, is computed from all particles belonging to the subhalo.

We define $F_S$ as the component of the ram force along the direction of gravity:
\begin{equation}
F_S = - \frac{\vec{F}_{\rm RAM} \cdot \vec{F}_{G}}{|\vec{F}_{G}|} \mbox{ .}
\end{equation}

By construction, $F_{S,j}$ can be positive or negative. If $F_{S,j} > 0$, the force acts against gravity ($F_{S^+,j}$); if $F_{S,j} < 0$, it acts in the same direction as gravity ($F_{S^-,j}$) (see Figure \ref{Fig:diagram}).

The total projected force is then
\begin{equation}
F_{S_{\rm Tot}} = \sum_{j=1}^{n} \left(F_{S^+,j} - F_{S^-,j}\right) \mbox{ ,}
\end{equation}
where $n$ is the number of gas particles in the subhalo. Using these quantities, we define a dimensionless \textit{RAMness} parameter:
\begin{equation}
{\rm RAMness} = \frac{F_{S^+}}{F_{S_{\rm Tot}}} \mbox{ ,}
\end{equation}
which ranges between $0$ and $1$. Values approaching $1$ indicate that ram pressure dominates over gravity, efficiently stripping gas and forming elongated features or tentacles, characteristic of jellyfish galaxies. Figure \ref{Fig:ramness_evol} shows snapshots of subhalos with increasing RAMness, confirming that the parameter robustly traces gas stripping.

Figures \ref{Fig:Ramness0} and \ref{Fig:Ramness1} illustrate the evolution of RAMness along the orbits of red subhalos, separated by infall time ($>6$ Gyr and $<6$ Gyr). Cluster-centric distances are plotted against look back time, with the infall time marked by a vertical dotted line and the host virial radius by a dark blue dashed line. Rainbow colors indicate the instantaneous RAMness, while red segments correspond to maxima, during which we constructed gas density maps weighted by $F_S/|\vec{F}_G|$ (bottom panels of Figures \ref{Fig:Ramness0} and \ref{Fig:Ramness1}), highlighting the stripped gas particles and visible tentacles.

Notably, red subhalos experience multiple jellyfish phases. Gas is not fully removed during the first peri-center passage, allowing subsequent episodes of gas stripping, sometimes persisting for $\sim7$ Gyr. This highlights that RPS can act over extended periods of time, producing repeated morphological features.

In a forthcoming paper (Gonzalez-Carbajal et al., in prep), we will report the RAMness calculation study across the entire TNG-50 sample, to produce a quantitative catalog of jellyfish galaxies and their gas tentacles.

% REFEREE COMMENT 4 (TIDAL)$
\textbf{It has to be noted that although tidal interactions may affect satellite evolution, their impact is expected to be primarily associated with changes in the stellar component and orbital heating. In contrast, the RAMness parameter is designed to quantify the direct competition between hydrodynamical stripping and the gravitational restoring force of the gas disk. Since the stellar rotational support of our galaxies remains approximately constant and subhalos with significant merger or flyby activity are excluded, we interpret the high-RAMness episodes as evidence that ram pressure is the dominant mechanism responsible for the observed gas depletion.}

%%%%%%%%%%%%%%%%%%%%%%
% Section Conclusions
%%%%%%%%%%%%%%%%%%%%%%
\section{Conclusions}
\label{sec:conclusions}
The transformation of blue spirals into early-type galaxies in low-mass clusters and group environments has traditionally been attributed to gravitational interactions, mergers, or gas exhaustion, typically occurring over $\gtrsim4$~Gyr after a merger event \citep{querejeta:17,deeley:21}. In dense clusters, ram-pressure stripping (RPS) provides an alternative, rapid pathway, quenching star formation and producing S0 galaxies within $\sim1-2$~Gyr after infall \citep{deeley:21}. Observationally, these processes are supported by the presence of passive spirals in groups \citep{cui:24}, as well as jellyfish galaxies in clusters displaying extended gas tails and localized star formation in stripped material \citep{poggianti:19,jones:22}.

% REFEREE COMMENT 1
\textbf{Using the IllustrisTNG-50 simulation, we investigated whether RPS can drive the formation of red spirals in satellite galaxies spanning environments from galaxy groups to low-mass clusters}. At $z=0$ we selected 41 subhalos with disky morphologies, low gas mass ($M_{\rm gas} < 5\times10^9,M_\odot$), and $M_* > 5\times10^9,M_\odot$, excluding any systems affected by mergers or flybys. Classifying galaxies via \textit{g-r} color into blue, green, and red subhalos \citep{cui:24}, we identified 11 red subhalos with infall times ranging from $\sim4$ to $7.3$~Gyr.
The gas and stellar mass histories of these red subhalos reveal a clear signature of RPS: rapid depletion of cold gas while the stellar component grows slowly, consistent with theoretical expectations for hydrodynamical quenching \citep{gunn:72}. Star formation is temporarily enhanced after infall due to gas compression, producing brief rejuvenated, bluer phases, before being ultimately quenched as gas is stripped. 

The transformation from blue to red spirals is slow, with an average timescale of $\sim6$~Gyr and up to $\sim7$~Gyr in some cases —longer than the rapid $\sim1-2$~Gyr quenching observed in massive clusters \citep{peng:10,vulcani:15,pasquali:19,galazzi:21}.

%REFEREE COMMENT 1
\textbf{Although RPS is generally expected to become more efficient in more massive halos, within our selected sample we find no significant dependence of the transformation timescale on host halo mass. Instead, the evolutionary histories indicate that orbital trajectories and cumulative exposure to ram pressure play the dominant role.}

Orbital histories reveal multiple peri-cluster passages, demonstrating that galaxies can experience several distinct jellyfish phases before losing all their gas (Figures~\ref{Fig:Ramness0} and \ref{Fig:Ramness1}). Using our \textit{RAMness} parameter (${\rm RAMness} = F_{S^+}/F_{S_{\rm Tot}}$), we quantify the hydrodynamical impact: high RAMness ($\gtrsim0.85$) coincides with elongated gas tentacles and active gas stripping, confirming RPS as the dominant transformation mechanism. These simulated signatures are consistent with observations of jellyfish galaxies in groups and clusters, exhibiting gas tails, and enhanced star formation \citep{poggianti:19}.

% REFEREE 4 (TIDES)
% ReWritten
\textbf{Our results indicate that RPS is the dominant environmental mechanism driving the transformation of the selected galaxies into red spirals. Because our sample was specifically constructed to minimize the effects of mergers, flybys, and other strong gravitational perturbations, the observed gas depletion and color evolution can be largely explained by hydrodynamical stripping. However, we cannot completely exclude weaker contributions from other environmental processes, including the tidal field of the host halo, starvation, viscous stripping, or thermal evaporation, which may operate simultaneously during satellite evolution. In this sense, our results complement previous studies in which environmentally driven tidal effects have been proposed as an important quenching mechanism for satellite galaxies \citep{weinmann:10}). More generally, the long-lived ram-pressure stripping events identified here are consistent with the broader picture in which RPS contributes to enriching the intracluster medium with metals, triggering star formation in stripped gas, and possibly promoting the formation of self-gravitating stellar systems within the stripped material \citep{jones:22,lora:24}.}

These findings have important theoretical and observational implications. They establish a natural evolutionary pathway connecting blue spirals to red spirals, and eventually to S0 galaxies, in low-mass clusters and group environments \citep{cui:24,masters:10}.
They provide testable predictions for the frequency, duration, and morphological signatures of RPS in groups, including repeated gas tails, moderate SFR enhancements, and long-lasting hydrodynamical effects that can be probed with integral-field surveys such as MaNGA and SAMI \citep{cui:24}.

In summary, this work highlights RPS as a slow but efficient mechanism shaping galaxy color, and star formation, over several Gyr. It complements gravitational and secular processes, showing that environmental quenching is a long-term, cumulative driver of galaxy evolution.

%%%%%%%%%%%%%%%%%%%%%
%\acknowledgments
%%%%%%%%%%%%%%%%%%%%%
\begin{acknowledgments}
We gratefully acknowledge the constructive comments and suggestions of the referee, which helped improve the clarity and quality of this manuscript.
VL gratefully acknowledges F.J. Sanchez-Salcedo for his comments and suggestions about this work.
VL gratefully acknowledges support from the SECIHTI Research Fellowship program. 
VL, IGC, JPO, and SED gratefully acknowledge support from the SECIHTI project CBF-2025-I-1764.
J.F acknowledges the support by PAPIIT-UNAM under grant IN102226.\\

\end{acknowledgments}

%%%%%%%%%%%%%%%%%%%%%
%%% Bibliography  %%%
%%%%%%%%%%%%%%%%%%%%%
\bibliographystyle{aasjournal}
\bibliography{references3}

%%%%%%%%%%%%%%%%%%%%%
%%%   Appendix    %%%
%%%%%%%%%%%%%%%%%%%%%
\appendix 
\renewcommand{\thefigure}{A\arabic{figure}}
\setcounter{figure}{0}
\section{RPS Vs Starvation}
We computed the fractional gas loss, and gas consumption due to star formation as a function of redshift. Gas loss occurs in strong, impulsive episodes correlated with pericenters and exceeds star-formation consumption, demonstrating that RPS dominates the gas depletion of our subhalo sample.

\begin{figure*}[h!]
    \centering
     {\includegraphics[width=0.28\textwidth]{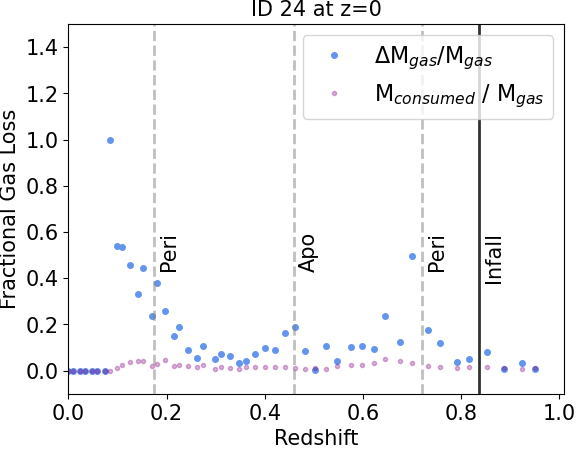}}
     {\includegraphics[width=0.28\textwidth]{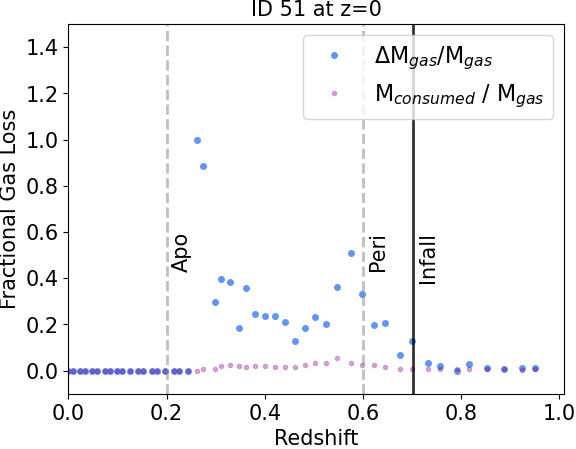}}
     {\includegraphics[width=0.28\textwidth]{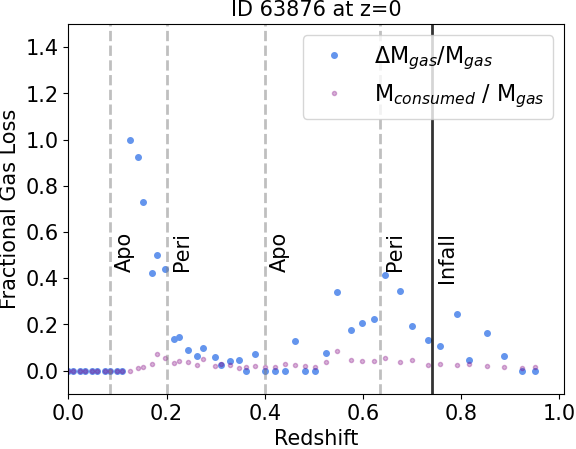}}\\
     {\includegraphics[width=0.28\textwidth]{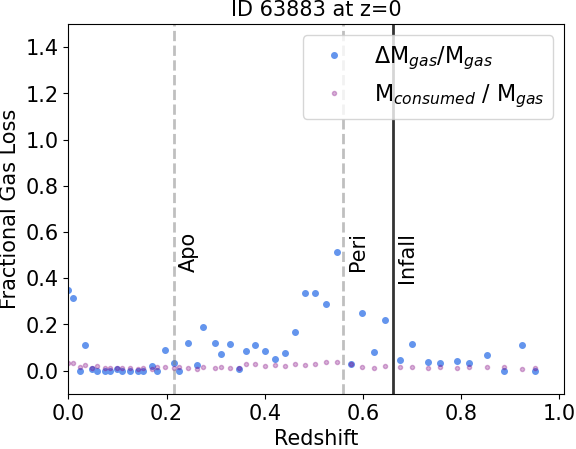}}
     {\includegraphics[width=0.28\textwidth]{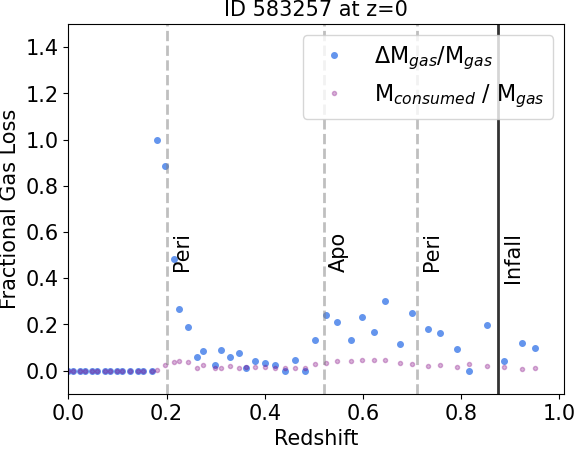}}   
    {\includegraphics[width=0.28\textwidth]{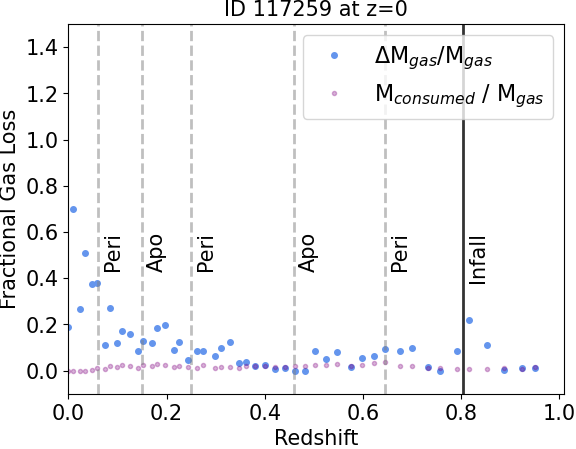}}\\
    {\includegraphics[width=0.28\textwidth]{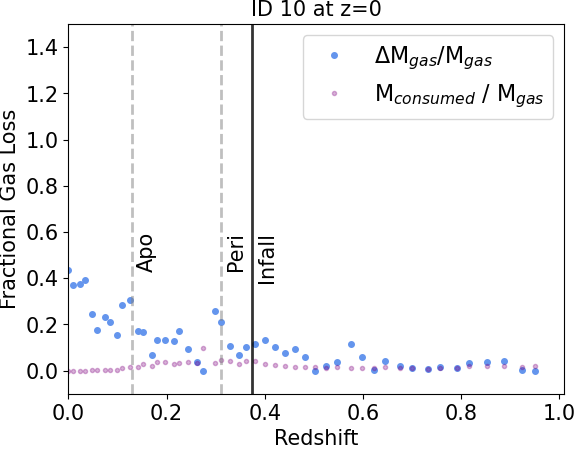}}
    {\includegraphics[width=0.28\textwidth]{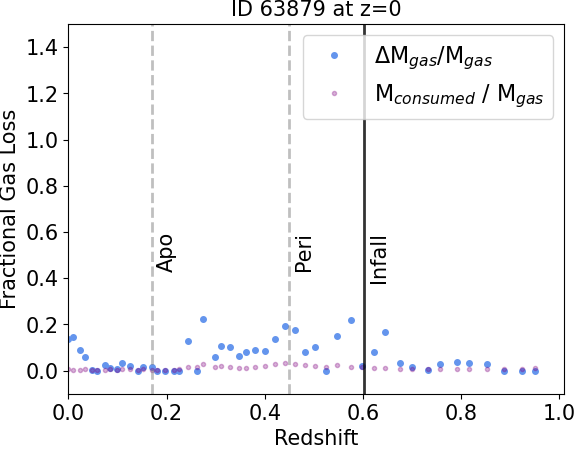}}
    {\includegraphics[width=0.28\textwidth]{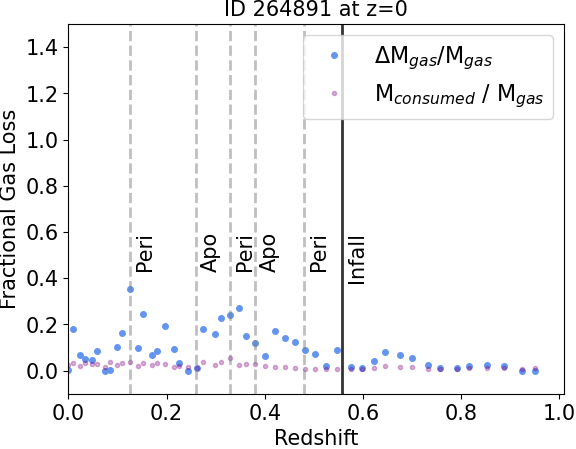}}\\
    {\includegraphics[width=0.28\textwidth]{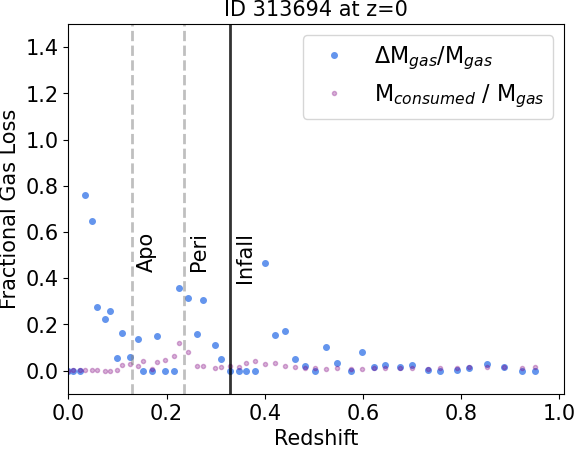}}
    {\includegraphics[width=0.28\textwidth]{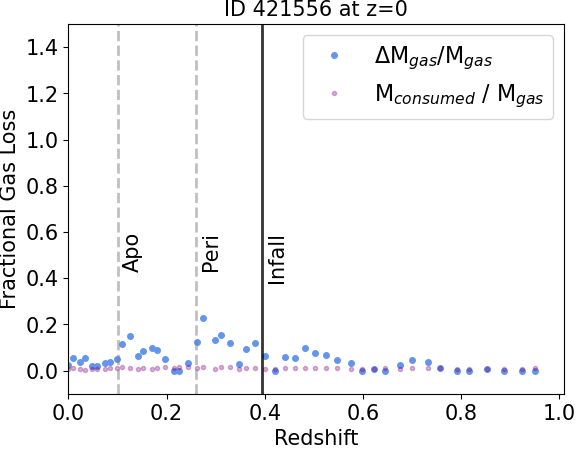}}
    \caption{Fractional gas loss versus star-formation consumption as a function of redshift. The blue points show the fractional change in bound gas mass per snapshot, while purple points show the fractional gas mass consumed by star formation, both quantities are normalized to the gas mass at the beginning of each snapshot interval. The vertical solid line, indicates the infall time. The vertical dashed lines indicate pericentric and apocentric passages.}     
    \label{Fig:appendix}
\end{figure*}

\end{document}